\documentclass[english,aps,prstper,reprint,showpacs,titlepage,longbibliography,floatfix,nofootinbib]{revtex4-2}   

\usepackage[T1]{fontenc}	
\usepackage[latin9]{inputenc}	
\usepackage{geometry}		
\usepackage{graphicx}
\usepackage{tabularx}
\usepackage{subfig}
\usepackage{amssymb}
\usepackage[above,below]{placeins}	
\usepackage{times}

\usepackage{hyperref}  
\hypersetup{colorlinks=true,urlcolor=blue,citecolor=blue,linkcolor=blue}   
\usepackage{enumitem}          
\setlist{nosep}                 
\begin{document}

\begin{titlepage}

  \title{A method to assess the trustworthiness of machine coding at scale}

  \author{Rebeckah K. Fussell}
  \author{Emily M. Stump}
  \author{N. G. Holmes}
  \affiliation{Laboratory of Atomic and Solid State Physics, Cornell University, Ithaca, New York 14853, USA} 


  \begin{abstract}
  Physics education researchers are interested in using the tools of machine learning and natural language processing to make quantitative claims from natural language and text data, such as open-ended responses to survey questions. The aspiration is that this form of machine coding may be more efficient and consistent than human coding, allowing much larger and broader data sets to be analyzed than is practical with human coders. Existing work that uses these tools, however, does not investigate norms that allow for trustworthy quantitative claims without full reliance on cross-checking with human coding, which defeats the purpose of using these automated tools. Here we propose a four-part method for making such claims with supervised natural language processing: evaluating a trained model, calculating statistical uncertainty, calculating systematic uncertainty from the trained algorithm, and calculating systematic uncertainty from novel data sources. We provide evidence for this method using data from two distinct short response survey questions with two distinct coding schemes. We also provide a real-world example of using these practices to machine code a data set unseen by human coders. We offer recommendations to guide physics education researchers who may use machine-coding methods in the future. 
  \end{abstract}

  \maketitle
\end{titlepage}

\section{Introduction}\label{Sec:introduction}
Education researchers are exploring the use of machine learning to automate the process of applying coding schemes to students' written work. Typically, human coders painstakingly apply coding schemes to such responses. ``Machine coding,'' by contrast, promises to utilize machine learning and natural language processing (NLP) tools to dramatically increase the efficiency of coding new data. With efficient machine coding, coding schemes could be applied at scale: across years, courses, and institutions. This scale would allow researchers to answer research questions and to evaluate diverse populations of students~\cite{Kanim2020} in ways that have been unavailable in the past because of the large amounts of manual coding that would be required. In addition to efficiency improvement, machine coding could improve consistency. Machine coding algorithms can be fixed such that they use the same procedure to code each response, whereas a human coder might code responses inconsistently because of fatigue or lack of clarity about the rules of the coding scheme. 

Skeptics of machine learning, however, worry that the algorithms will introduce or perpetuate biases in training data~\cite{burtonSystematic2020} or distrust that an imperfect algorithm may be preferable to imperfect human judgment~\cite{dietvorstOvercoming2018}. Here, we seek to address this skepticism by presenting methods of evaluating the trustworthiness of a machine learning algorithm specific to physics education research contexts, drawing on data analysis techniques common to experimental physics (namely, quantifying statistical and systematic uncertainties). 

We argue that these techniques particularly support physics education researchers because we cannot necessarily rely on mainstream machine learning techniques. For instance, one common NLP exercise is to train and test algorithms using aggregated banks of news headlines,~\cite[e.g.,][]{misra2021sculpting,misra2022news}, which contain many thousands of unique instances of written text and come pre-labeled by news curators with codes like ``politics'' and ``wellness''. 
By contrast, education data sets of student writing are often small (e.g. one thousand or fewer short paragraph responses) and training data generally would need to be labeled by researchers. 

In this work, we focus on machine learning processes that can be designed to mimic human coding of students' written text (responses). These processes center around the coder learning a set of rules (a ``trained model'' in machine learning or a ``coding scheme'' in human coding) that define whether or not a label should be applied to a type of response. The coder then applies labels to responses based on these rules. Supervised machine learning algorithms are particularly aligned to mimic human coding. A supervised algorithm uses a training set of human-coded data to learn the set of rules, then applies these rules to machine code any new data shown to the trained algorithm. 

Machine coding with a supervised algorithm in this manner is an emerging research practice in physics education research, not simply a technical procedure. Machine coding is a practice that aims to make quantitative measurements and claims about student responses, such as measuring the frequency of categories of ideas and themes in students' written responses. We contrast these quantitative measurements and claims to qualitative measurements and claims, such as identifying the existence of ideas and themes. As a new quantitative research practice, it is vital in these early stages to establish high-quality normative practices~\cite{Ding}. 



Unfortunately, physics education research has not yet established these norms, such that we can apply machine coding to student text at scale, which we relate to two key issues. First, relatively few studies have focused on using machine learning for making quantitative claims (for example, claims about the quantity or frequency of codes within a data set). Instead, most of the existing work by education researchers that utilizes machine learning of written text focuses on making qualitative claims. For example, researchers have used unsupervised techniques to identify themes within text and video data and to generate insights that complement human analytical insights~\cite{Sherin2013,Sherin2018,Hur_etal_2023,Kubsch_Krist_Rosenberg, Mariegaard_Seidelin_Bruun}. Others have performed computational grounded theory to aid the human analyst in developing theoretical constructs from text data~\cite{Rosenberg_Krist,Nelson2020, Tschisgale_Wulff_Kubsch}. 

Second, the studies that provide ``proof of concept'' towards automating coding to make quantitative claims at scale~\cite[e.g.,][]{Wilson_etal_2022,Nakamura_etal,Fussell_etal,Campbell_etal, Ullmann, Wulff_etal, MattThomas,Ruijie,Nelson_etal_2021,Atteveldt_etal} primarily rely on human-coded data to test the validity of their models and to establish trust in any quantitative claims. For example, researchers using thematic analysis (sometimes called Topic Modeling in the natural language processing field) to evaluate the prevalence of various themes within text data~\cite{Odden_etal_2020,Odden_etal_2021, Geiger_etal, Bralin_etal} noted that, due to the unsupervised nature of their form of analysis, these prevalence values cannot be taken to be a measurement of the ``ground truth'' as there is no way to assess the accuracy without comparing to human coders~\cite{Odden_etal_2020}. 
Researchers using supervised algorithms commonly evaluate the validity of machine coding by computing reliability metrics that compare the codes assigned by the algorithm to the codes assigned by a human. Common reliability metrics include Cohen's kappa, Quadratic Weighted Kappa, accuracy, recall, precision, F1 score, and area under the Receiver Operating Characteristics curve (AUC - ROC)~\cite{Rosenberg_Krist,Wilson_etal_2022, Nakamura_etal,Fussell_etal, Campbell_etal}. When calculating these reliability metrics, researchers need a large human-coded data set that can be split into sufficiently large training and test sets. 

Even then, while researchers have developed algorithms that can surpass threshold values for reliability~\cite[e.g.,][]{Rosenberg_Krist,Wilson_etal_2022,Fussell_etal,MattThomas}, it is not clear what to make of these reliability measures when it comes time to make quantitative education claims about machine coded datasets. For example, if a machine coding model obtains a Cohen's kappa value of 0.65, indicating ``substantial agreement'' with a human coder~\cite{LandisKoch1977}, we are left with questions such as: did the computer systematically over- or under-estimate the code relative to the human coder? If we make a quantitative education claim using the machine coded data, how much uncertainty should we maintain toward that claim given this value of Cohen's kappa?

Furthermore, there is no guarantee that the threshold will continue to be met in other, novel data sets. The field does not yet have established norms for evaluating the validity of the coding in novel data sets. 
This concern is especially relevant when the data sets being assessed are small or systematically different from the training dataset, as in most human-coded education datasets. For example, the reliability metric may change significantly if the algorithm were applied to a test set from a new institution or student population. Researchers in computer science and statistics are developing advanced methods to tackle this problem by improving the ability of trained models to accurately process test data that is unlike the training data using a causal inference perspective~\cite{Wang_etal} or by leveraging the causal instincts of human annotators~\cite{Srivastava_etal}. These methods, however, require substantial resources (e.g., amount of data and amount of human-coding) beyond what is accessible to physics education researchers.


For now, as physics education researchers rely on the reliability metric approach to evaluating machine coding, we are left with a dilemma: if the only way to evaluate the accuracy and reliability of machine coding is to human-code all (or at least a very large fraction) of the data, then what is the benefit of machine coding?

Here we propose a set of methodological practices that can be used to evaluate the trustworthiness (accuracy and reliability) of machine coded data to make quantitative education research claims while minimizing the need for human-coding through uncertainty quantification. Researchers in other fields have similarly analyzed the quantification of uncertainty when using natural language processing~\cite[e.g.,][]{hu2023uncertainty}, but these methods do not provide specific steps that allow researchers in our field to make and evaluate trustworthy quantitative claims. Our goal is to contribute to the conversation on methodological standards and review of quantitative claims within PER~\cite{Odden_Fussell_Green_Young_PERC22, Adlakha_Kuo}, and we encourage others in the community to expand on our framework.

We describe a four-part methodology for evaluating machine coding that casts machine coding results as experimental measurements with associated uncertainties (statistical and systematic). The methodology: i) evaluates a trained model, ii) quantifies statistical uncertainty, iii) quantifies systematic uncertainty from the trained algorithm, and iv) quantifies systematic uncertainty in novel data sets. We argue this approach addresses the concerns described above and incorporates best practices in expressing measurements with uncertainty~\cite{GUM}. 

The rest of this paper is organized as follows. First, in the methods section we present our data sources and our approach to training a supervised machine learning algorithm that uses natural language processing to perform machine coding. Then, we devote a section to each part of the methodology. We then work through an example of applying this methodology to a data set we have not coded by hand. Lastly, we discuss the limitations of the evidence available to us so far and suggest opportunities for future work.

\section{Methods}\label{Sec:methods}
\subsection{Data sources}
In this work, we use two distinct sets of student responses to open-response survey questions: the \textit{Trustworthy} data and the \textit{Sources} data. 

The Trustworthy data consists of 1,958 responses to the survey question ``How do you know whether or not an experimental result is acceptable or trustworthy? What
gives you confidence that the data is trustworthy?'' from Ref.~\cite{hu_zwickl}. The question was included on pre- and post-surveys administered at Cornell University across three different academic years. Data are from students taking calculus-based introductory mechanics and electricity and magnetism courses. We developed a coding scheme with seven codes adapted from the scheme used in Ref.~\cite{hu_zwickl}: Consistent Results, Uncertainty, Expected Result, Good Methods, Ethics, Peer Review, and Statistics. In this coding scheme, individual responses can receive multiple codes and the coders identify the presence or absence of each code one at a time. 

In the main analysis below, we focus on one of these codes: Consistent Results (CR). For the CR code, two human coders achieved a Cohen's kappa value of 0.9 in 10\% of the data. A kappa value of 0.8-1.0 is considered to be very good agreement~\cite{LandisKoch1977}. Then one human coder hand-coded 1,672 of the responses -- all the data from semesters prior to Fall 2022. The remaining 286 responses were from the end of the Fall 2022 semester and were not coded by humans. 

The CR code is defined as: ``the same result is obtained when looking at more data, either through repeated trials or comparing with other people.''  The inclusion criteria for the CR code are: 
\begin{enumerate}
    \item repeated, repeatable, and/or consistent across independent measurements, 
    \item experiment conducted numerous times, 
    \item obtaining consistent results, 
    \item repeating experiments with different methods, 
    \item collecting additional data to verify trends, 
    \item multiple trials, 
    \item low variance or deviation from mean, 
    \item large sample size, 
    \item comparing with peers, other research groups, different scientists, published results, 
    \item replicability or reproducibility, 
    \item consistency of results across different research groups, 
    \item others get the same results.
\end{enumerate}

In Sec.~\ref{Sec:worked_example}, we also examine another of these codes: the Uncertainty code, which will be defined there.

The Sources data consists of 2,413 responses to a survey where respondents were shown experimental physics scenarios and fictional distributions of data. The responses in this analysis are to the question ``What is causing the shape of the distribution measured by the students? List as many causes as you can think of.'' from Ref.~\cite{Stump_etal_sources}. The survey was administered in over a dozen courses at 12 institutions. In total, the responses were written by 753 students, each of whom wrote multiple sources for each experimental scenario (each source is treated as a unique response). Upper-division students were shown up to four different experiments and prompted to list sources of uncertainty for each, so the data set consists of students' ideas about uncertainty for projectile motion (468 responses), Brownian motion (290 responses), single slit (239 responses), and Stern-Gerlach (282 responses) experiments~\cite{Stump_etal_sources}. An additional 1,134 responses for the projectile motion scenario came from introductory-level students~\cite{Stump_etal_focused_collection}. The responses were coded based on three categories -- Limitations, Principles, and Other -- as described in Ref.~\cite{Stump_etal_sources}. 
Two human coders achieved a Cohen's kappa of 0.85 across 40 responses. Then each coder coded a portion of the rest of the data. 

In the analysis below, we focus on the Limitations code (L), which was by far the most frequent code. The L code is defined as: ``A distribution is caused by practical limitations in an experiment owing to our inability to perfectly model and measure a real-world system.'' No inclusion criteria are listed in the coding scheme. Instead, human coders primarily used exclusion criteria in coding. The exclusion criteria for the L code are: 
\begin{enumerate}
    \item A distribution is caused by some principle of theoretical physics (theoretical abstraction of an experiment e.g. quantum theory) or of experimentation (measurement is inherently random/ has an inherent normal distribution). 
    \item Vaguely worded responses about ``uncertainty'' or ``random errors''. 
    \item Physical mechanisms that determine the position (distance) of the central value (average value) but are not varying between experimental trials, e.g. gravity.
\end{enumerate}
We chose to use these two codes because they are reliably measured by human coders and provide one example of a code mostly defined by inclusion criteria (CR) and one example of a code mostly defined by exclusion criteria (L). This choice allows us to investigate the applicability of these methods across a range of rule types that physics education researchers may encounter.

\subsection{Natural Language Processing}

In developing the machine coding algorithms, we used a one-vs-all (OVA) approach, where we built separate algorithms that focused on one code at a time. The machine applies a label of $0$ or $1$ if the code is absent or present, respectively. The OVA approach mirrors the decision-making process the human coders used (namely, focusing on one code at a time and reading each response for evidence of that code) and has been shown to perform as well as more complex, multi-code approaches~\cite{Rifkin_Klautau}. 

To prepare the responses for automated coding, we used the following bag-of-words natural language processing (NLP) protocol: i) split all words in each response into individual words (often called tokens or features), ii) fix contractions (for example, ``you're'' becomes ``you are''), iii) use the Word Net Lemmatizer from the Natural Language Toolkit python package~\cite{nltk} to combine words from the same family such as plurals and verb conjugates, and iv) remove any remaining whitespace, punctuation, and numbers. We then encode the modified responses as a matrix where each row corresponds to a single response and each column represents a single token (i.e., each unique word in the total set of modified responses). Each entry in the matrix is a 0 or 1 indicating whether that token is present in the response. We found in initial testing that this 0 or 1 method outperforms other common word scoring methods such as raw count of each word or Term Frequency Inverse Document Frequency~\cite{tfidf, brownlee}.

We did not filter out stop words such as ``and'', ``it'', and ``the''. This form of filtering can be useful to reduce the size of a large data set or to perform machine coding tasks such as categorizing technical texts~\cite{Sarica_Luo}. 
In an education research context, however, we are dealing with small data sets where reducing the size of the data in pre-processing may be a less important consideration. Furthermore, stop words like ``only'', ``but'', and ``just'' may be critical when analyzing the nuances of student thinking as revealed by their responses~\cite{Tannen}.

In this paper, all machine learning models were built with bag-of-words logistic regression algorithms using the scikit-learn package in python~\cite{scikit-learn}. In initial testing, we examined bag-of-words models built with Logistic Regression, Naive Bayes, SVM, Random Forest, k-Nearest neighbors, and neural networks (both a single-layer perceptron and a convolutional neural network utilizing the GloVe word embedding~\cite{keras, glove}). We found that Logistic Regression and single-layer perceptron had the highest, most consistent performance. We chose to use Logistic Regression because it is easier to view the coefficients assigned to different features and because it has been used in other physics education research literature, such as \cite{Wilson_etal_2022}. 

\subsection{Training and test sets}
\begin{figure}{\includegraphics[width=0.9\linewidth]{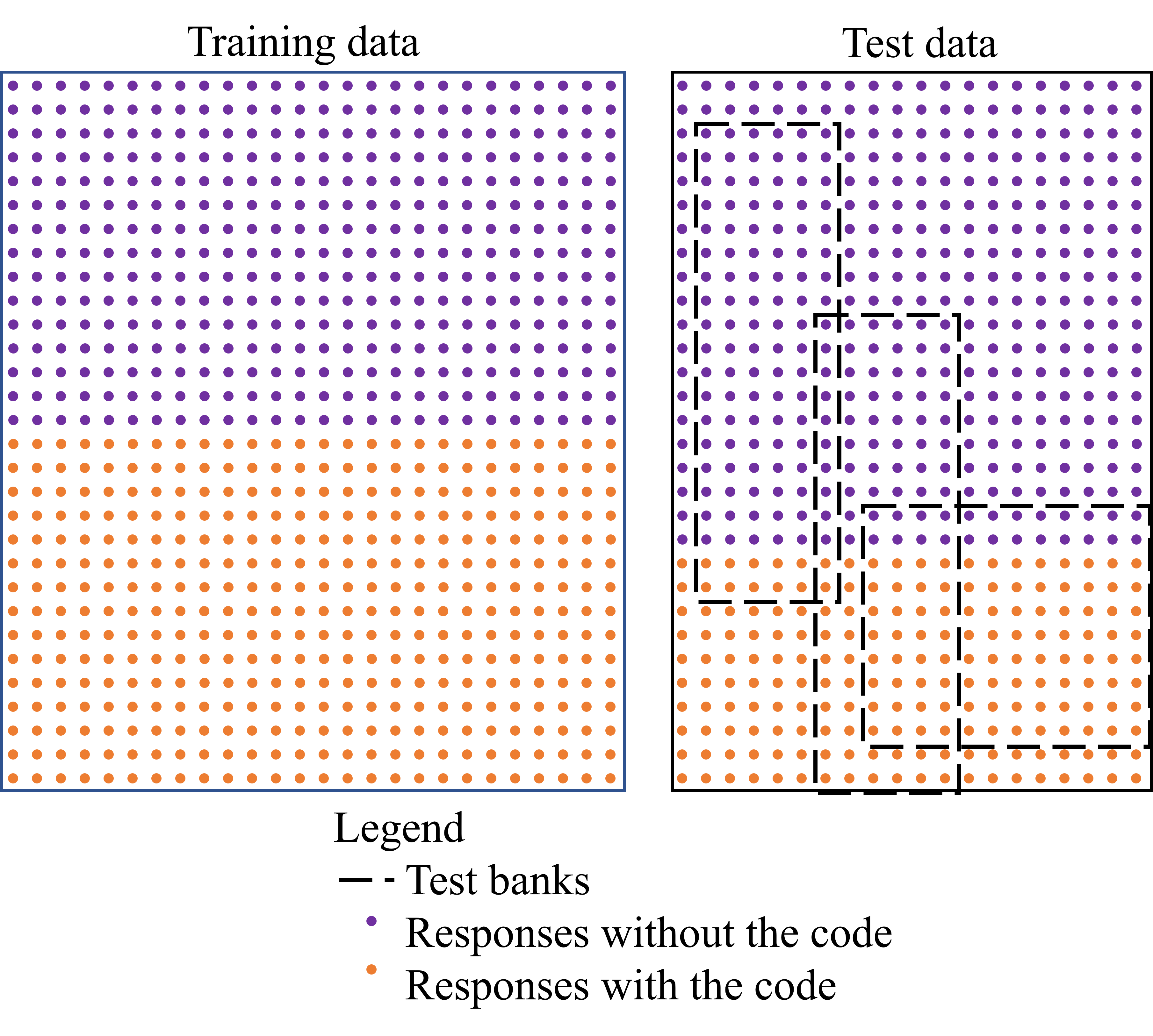}}
  \caption{The available data are split into training data and test data. Purple points represent responses without the code and orange points represent responses with the code. In this representation, three test banks of fixed size $N_{bank}$ are drawn outlined in a dashed line, each with a different fixed proportion of responses that contain the code: 0.2, 0.5, and 0.8. Test set samples of size $n$ are drawn at random from the test banks (in the figure, data are sorted so random samples are not pictured). ~\label{data_slicing}}\vspace{-1em}
\end{figure}

We split the available responses to create a training data set and a test data set (Fig.~\ref{data_slicing}); details about the test data are further described in Section~\ref{Subsec:test_set}. In the majority of our analyses, the training data includes $N_{train} = 600$ responses to train each algorithm. While algorithm performance generally increases as the size of the training set increases, we previously found that performance plateaus around $N_{train} = 600$ student responses from the Trustworthy data~\cite{Fussell_etal}. In a few analyses, we use a smaller training set because we need additional data for testing. We note these instances as appropriate. We define $p_{train}$ as the proportion of responses in the training set containing the particular code, where $0 \leq p_{train} \leq 1$. 

During model development, we keep the training data and test data silo-ed from each other as suggested by Aiken et al~\cite{Aiken_etal_2021}. 
Overestimation of model performance can occur when training and test data are repeatedly split at random from the full data set and not kept strictly separate during model development. This contamination can occur even if test data are not directly included in the training set (for instance, information from the test set can enter training during feature selection or hyperparameter optimization), especially when sample sizes are small~\cite{Vabalas_etal}. By keeping the two sets silo-ed, we remove this risk. 

Whereas, in previous work, researchers reported results averaged across multiple training and test sets~\cite[e.g.,][]{Fussell_etal, Wilson_etal_2022, Nakamura_etal,Young_Caballero}, here we analyze only one training set at a time, though we use multiple sample test sets (each drawn from a test bank). We suggest that this process better mimics the intended process for education research in which one trained algorithm would be used to machine code new data. In addition, with limited resources, a researcher should prioritize human-coding one large training set as opposed to many smaller training sets, because a larger training set tends to create a more reliable algorithm~\cite{Fussell_etal}. 

\subsection{Test Set Sampling}\label{Subsec:test_set}

We divide the test data into a set of test banks, each of which is of size $N_{bank}$. In the analysis below, we intentionally create test banks that span the range of possible proportions of responses that contain the code (as determined by human coders) between 0 and 1 inclusive, usually a multiple of $0.1$. 
We then draw multiple sample test sets of size $n$ from each test bank upon which we perform our analyses. Naturally, $N_{bank} > n$ and we seek to select values of $n$ such that $N_{bank}$ is 2-3 times larger than $n$. This number allows us to balance the need to produce samples that are different from each other with the constraints of limited data. 

The trained model computes the proportion of responses in the sample test set that contain the code, which we call $E_{C}$ (the ``computer's estimate''). We re-sample the test bank a set number of times (usually 100 or 1000) to develop a distribution of $E_C$ and to calculate $\overline{E_C}$, the mean of the distribution of repeated measurements of $E_{C}$ made with the different sample test sets.

We define $E_H$ as the proportion of responses in \emph{each sample test set} containing the code as determined by the human coder. With enough sample test sets pulled from the test bank (that is, when the number of re-samplings of the test bank is sufficiently large), the mean of $E_H$ across the sample test sets, $\overline{E_H}$, is approximately equal to the proportion of responses in the test bank containing the code, based on statistical properties of sampling from a population.

\subsection{Optimizing a training model}\label{optimizing}
A machine coding algorithm can be optimized in multiple ways during development and researchers must make informed decisions about optimization that are relevant for their particular needs. For discussion of optimizing hyperparameters (fixed parameters that determine how the algorithm behaves) in logistic regression models, we refer the reader to~\cite{Wilson_etal_2022}.

In this article, we focus discussion about optimization on a practical consideration that is particularly relevant to physics education researchers, namely using training data that has an equal distribution of responses across each of the possible outcomes (referred to as ``outcome balance''). For example, when using one-vs-all coding as in this work, a balanced training set would have 50\% of the responses in the training data set with the code (i.e., $p_{train} = 0.5$). We focus on this consideration as one particularly relevant to education researchers who are constrained by the size of educational data sets and the necessary investment of human coding for developing training sets.


\begin{figure}{\includegraphics[width=0.9\linewidth]{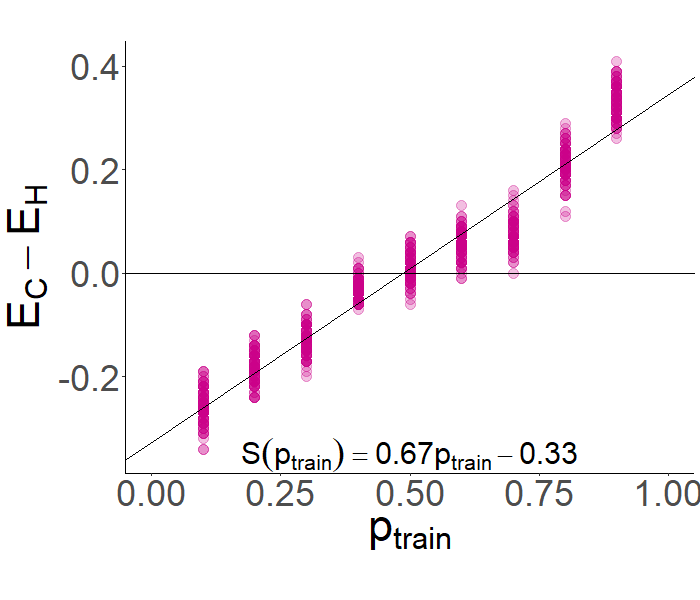}}
  \caption{The difference between the human ($E_H$) and computer's ($E_C$) estimate of the presence of a code in a test set increases as $p_{train}$ is less balanced (i.e., far from $p_{train} = 0.5$). (Code: CR, proportion of responses in the test bank containing the code is $0.5$).~\label{fix_test_vary_train}}\vspace{-1em}
\end{figure}

This recommendation is motivated by previous work, 
such as Refs.~\cite{Fussell_etal, Young_Caballero}, which suggest that an algorithm becomes less reliable when trained with a data set with unbalanced outcomes (that is, a code is present in a majority or minority of responses). 
The benefits of balanced representation of the presence and absence of a code also aligns with models of human learning, such as through contrasting cases~\cite{Bransford_Schwartz} and negative instances or non-examples~\cite{Shumway}.


We demonstrate the impact of balancing the training set by training our algorithm with nine different training sets for our CR code, each with a different value of $p_{train}$ between $0.1$ and $0.9$ inclusive. The size of the training set for each algorithm was $600$ responses. We then create a set of test banks where each test bank is of size $N_{bank} = 200$ responses, each test bank is sampled 100 times, and each sample test set is of size $n = 100$ responses. For each algorithm, we compute the proportion of responses with the code according to the computer coder, $E_C$, and according to the human coder, $E_H$. 
We fix the proportion of responses in the test bank containing the code (according to human coders) to be $0.5$. 

We observe that the magnitude of the difference between $E_C$ and $E_H$ deviates from zero as $p_{train}$ deviates from $0.5$ (Fig. \ref{fix_test_vary_train}). 
The difference is such that the machine coding overestimates the prevalence of the code in the test set (i.e., $E_C > E_H$) when $p_{train} > 0.5$ and the machine coding underestimates the code (i.e., $E_C < E_H$) when $p_{train} < 0.5$, with the effect increasing as $p_{train}$ becomes more unbalanced (farther from $0.5$). Crucially, we find no offset between the machine coding and the human coding when $p_{train} = 0.5$. In our supplemental materials, we show that these findings also hold for the L code from the Sources data set (Fig. \ref{fix_test_vary_train_L}). 

An important question is whether the lack of offset at $p_{train} = 0.5$ is based on a core principle of learning or simply that the offset is zero if $p_{train}$ equals the proportion of responses in the test bank (which in this case was set to $0.5$). We test this effect empirically in the supplemental materials for two different test set proportions. We indeed find that the algorithm better matches human coding when the proportion of responses with the code in the test set matches the value of $p_{train}$. In educational research settings, however, our goal is to apply a trained model to new data that has not been coded by humans, such that the proportion of responses containing the code is unknown. The analyses in the supplemental materials indicate that using training data with $p_{train} = 0.5$ is the best option to use across a range of possible test set proportions and, thus, is the best option for applying training models to new data (Fig.~\ref{appendix_other_test_prop}).

Thus, we propose balancing the training set as a key consideration when optimizing a trained model and we use balanced training sets throughout this paper. This consideration places constraints on the idea in previous work that increasing the size of a training set will improve the model~\cite{Fussell_etal}. A smaller, balanced training set may be better than a larger, unbalanced training set. 
We generally recommend that some balancing method (whether it be balanced training sets or a different balancing procedure such as random over-sampling of the minority case~\cite{Batista_etal}) be used in future PER research involving machine coding. 

One consideration for many researchers is: do I have enough human-coded data to do this type of analysis? We use 600 human-coded responses in the training set, as justified above. To balance the 600 responses in the training set, it is generally necessary for human coders to process more than 600 responses in order to find 300 responses that contain the code and 300 responses that do not. In some cases, if the code is well-captured by the trained algorithm, a smaller training set may be suitable. An example of such a case can be seen in Sec.~\ref{Sec:worked_example}.

The methods described up to this point leave us with a single trained model - a ``machine coder'' - that has learned a set of rules to consistently apply to each response. For more methodological details, see the code for this research on GitHub~\cite{github}.

\section{Evaluating a trained model}\label{Sec:evaluating}

The first step in establishing the trustworthiness of an algorithm is to check whether the trained algorithm has conceptual or theoretical validity. This is typically carried out by evaluating the outputs of the algorithm on the training set to see how it uses and connects individual words to codes~\cite{Sherin2013, Nelson2020}. We can also peek at the coefficients the model applies to individual words or other features. We think of this as ``looking under the hood'' of the algorithm to understand how it is characterizing words and codes.

\subsection{Explanation of method}

After optimizing the algorithm with a balanced training set, researchers can evaluate, qualitatively, which linguistic features the trained model uses to decide whether to include or exclude a response from a code. 
We identify these linguistic markers by evaluating the coefficients the model uses to process individual words. These coefficients are provided through the algorithm's logistic regression model, which models the probability $p$ that a response contains the code as a sigmoid function of the features, $x_i$ (in this case individual words), defined as:
\begin{equation}
    p = \frac{1}{1 + e^{-z}}
    \label{eq:logreg}
\end{equation}
where $z = \sum_{i} \beta_i x_i$ and $\beta_i$ are the coefficients that best fit the training data. For each response, $x_i = 1$ if the word is present and $x_i = 0$ if the word is absent. In the machine learning process, a regularization procedure is applied that shrinks the coefficients $\beta_i$ proportional to the square of the coefficient, making this a ridge regression. We denote the new coefficients (after shrinking) as $\kappa_i$. The shrinking procedure protects against overfitting by preventing the coefficients from growing too large. Note that the features $x_i$ in this case are individual words from the training set.

The coefficient $\kappa_i$ gives us information about if the algorithm uses that word as an inclusion or exclusion criteria for characterizing the code. A word with a coefficient greater than zero indicates the machine coding interprets the use of that word as evidence for \emph{inclusion} of the response in the code, while a word with a coefficient less than zero indicates the machine coding interprets the use of that word as evidence for \emph{exclusion} of the response from the code. The larger the magnitude of a coefficient, the more impact the corresponding word will have on the trained model. We then evaluate the extent to which these coefficients align with the coding scheme. A lack of alignment implies that human and machine coders use fundamentally different linguistic coding rules. Next, we provide an example of comparing the highest and lowest coefficients to the inclusion and exclusion criteria in a human-generated coding scheme. 

\subsection{Evidence and Examples}
We examine the features (individual words, $x_i$) of our two trained models (one for the CR code and one for the L code) along with their associated coefficients, $\kappa_i$. In Tables \ref{tab:CR_main} and \ref{tab:PLO} we show the words with the highest and lowest coefficients for the trained model for each coding scheme, which we compare to the inclusion and exclusion criteria in the associated coding schemes (described in Sec.~\ref{Sec:methods}).

\begin{table}[!htbp]
  \caption{Coefficients for the trained model for the CR code. Left side (coefficient > 0) corresponds to inclusion criteria in the trained model and right side (coefficient < 0) corresponds to exclusion criteria. Words in bold are listed in the inclusion criteria for the human-generated coding scheme for this code (or share a root and are part of the same word family as a word in the inclusion criteria, e.g. ``repetition'' and ``repeating'').~\label{tab:CR_main}}
  \begin{ruledtabular}
    \begin{tabularx}{\textwidth}{c c | c c }
      \textbf{Word} & \textbf{Coefficient} & \textbf{Word} & \textbf{Coefficient} \\ 
      \hline
      \textbf{trial} & 3.01 & analysis & -1.09\\
      \textbf{repeatable} & 2.23 & accurate & -0.92 \\
      \textbf{repeated} & 1.90 & how & -0.90 \\
      \textbf{multiple} & 1.80 & model & -0.72\\
      \textbf{replicable} & 1.73 & bias & -0.72\\
      \textbf{same} & 1.51 & uncertainty & -0.71\\
      lot & 1.47 & unbiased & -0.70\\
      similar & 1.32 & statistical & -0.68\\
      \textbf{time} & 1.30 & theoretical & -0.66\\
      \textbf{repetition} & 1.23 & reviewed & -0.62\\
      \textbf{consistent} & 1.09 & look & -0.61\\
      \textbf{replicability} & 1.08 & possible & -0.56\\
      \textbf{reproducible} & 1.06 & margin & -0.55\\
      many & 1.04 & creating & -0.54\\
      \textbf{others} & 1.02 & fair & -0.54\\
      \textbf{replicated} & 1.00 & need & -0.53\\
      \textbf{repeatability} & 0.99 & into & -0.52\\
      \textbf{reproducibility} & 0.83 & properly & -0.52\\
      \textbf{result} & 0.82 & wa & -0.51\\
      whether & 0.81 & t & -0.49\\
    \end{tabularx}
  \end{ruledtabular}\vspace{-1em}
\end{table}

For the CR code (Table~\ref{tab:CR_main}), the words with the highest coefficients are all either represented in the inclusion criteria or are close synonyms of words in the inclusion criteria of the human-generated coding scheme (for example, a ``lot'' is close to ``multiple'', ``similar'' is close to ``consistent''). Three of the twelve coding scheme inclusion criteria (1, 7, and 8) are not represented in the list of high coefficients. These three inclusion criteria are less common in the data and may be missed by machine coding. Looking at the distribution of coefficients, the trained model relies more on positive coefficient words compared to negative coefficient words (the magnitude of the positive coefficients are much larger than the magnitude of the negative coefficients), which supports the way human coders relied more on inclusion, rather than exclusion, criteria when coding for CR. 

\begin{table}[!htbp]
  \caption{Coefficients for the trained model for the L code. Left side (coefficient > 0) corresponds to inclusion criteria and right side (coefficient < 0) corresponds to exclusion criteria. Words in bold are listed in the exclusion criteria for the human-generated coding scheme for this code (or share a root and are part of the same word family as a word in the exclusion criteria, e.g. ``random'' and ``randomness'') \label{tab:PLO}}
  \begin{ruledtabular}
    \begin{tabularx}{\textwidth}{c c | c c }
    \multicolumn{4}{c}{}\\
 
      \textbf{Word} & \textbf{Coefficient} & \textbf{Word} & \textbf{Coefficient} \\ 
      \hline
      human & 2.49 & \textbf{distribution} & -1.41\\
      friction & 2.07 & \textbf{randomness} & -1.15\\
      measuring & 1.56 & \textbf{gravity} & -1.08\\
      not & 1.51 & \textbf{distance} & -1.05\\
      ruler & 1.49 & variation & -1.01\\
      ball & 1.38 & motion & -0.97\\
      different & 1.25 & statistical & -0.96 \\
      table & 1.22 & off & -0.95\\
      source & 1.10 & density & -0.90\\
      ramp & 1.06 & \textbf{quantum} & -0.90\\
      resistance & 1.05 & would & -0.89\\
      in & 1.05 & because & -0.89 \\
      material & 1.03 & \textbf{value} & -0.84\\
      condition & 1.03 & count & -0.81\\
      between & 0.95 & \textbf{random} & -0.79\\
      difference & 0.94 & data & -0.79\\
      incorrect & 0.93 & probability & -0.75\\
      slightly & 0.92 & answer & -0.75\\
      wind & 0.89 & \textbf{experimental} & -0.75\\
      inconsistency & 0.86 & each & -0.74\\
      initial & 0.83 & \textbf{average} & -0.74\\
      too & 0.82 & \textbf{principle} & -0.73\\
    \end{tabularx}
  \end{ruledtabular}\vspace{-1em}
\end{table} 

For the L code (Table~\ref{tab:PLO}), the magnitudes of the negative coefficient words are much larger than we saw for the CR code.
All exclusion criteria from the human-generated coding scheme are also represented by the words with low coefficients. This result again matches the human coding scheme, which relies almost exclusively on exclusion criteria, rather than inclusion criteria. Based on the list of words, two of the exclusion criteria (``vaguely worded answers'' and ``physical mechanisms that are not varying between experimental trials'') are not obviously captured by the algorithm. These criteria, however, by definition do not have defining words, and all the examples explicitly listed in these exclusion criteria (e.g. gravity) are present in the low coefficient words. 

\subsection{Recommendations}

We identify three key recommendations for evaluating a trained algorithm:
\begin{enumerate}
\item Evaluate the correspondence between the inclusion criteria of the coding scheme and the high (positive) coefficient words. Do all the words with a high coefficient correspond to an inclusion criteria? Are all inclusion criteria represented in the high coefficient list? 
\item Evaluate the correspondence between the exclusion criteria of the coding scheme and the low (negative) coefficient words. Do words with low coefficients correspond to exclusion criteria? Are all exclusion criteria represented in the low coefficient list?
\item Compare the relative distribution of positive coefficients to the distribution of negative coefficients. Are the relative magnitudes appropriate based on the importance of inclusion versus exclusion criteria in the coding scheme and the answers to the previous questions?

\end{enumerate}

If the answer to any of the questions is no, an element of the coding scheme may not be captured in the machine coding. In some cases, such as coding schemes with inclusion or exclusion criteria that are rare in the data, this misalignment may be acceptable as long as the researcher is aware of the omission when interpreting results from machine coding. If the researcher thinks this misalignment is unacceptable, consider expanding the size of the training set to include more examples of underrepresented criteria.

\section{Statistical Uncertainty}\label{Sec:statistical}

The second step in establishing the trustworthiness of an algorithm is to understand the statistical variability associated with applying the algorithm to a finite number of samples drawn from a larger population. Whenever we calculate the frequency of a code in a finite data set, we are sampling only a subset of the total population. If we were to take repeated measurements of the frequency of a code in multiple samples drawn from the broader student population, we would find natural variation across the measurements. This variation occurs because each of the finite sample of responses features a degree of random variation. The random variation in these samples leads to variation in the frequency a single coder (whether human or machine) would measure between samples. 
The natural variation caused by these factors can be quantified by calculating the standard deviation of many randomly selected samples, applying principles of statistics to machine coding student data. 

\begin{figure*}[t]
  \subfloat[]{\includegraphics[width=0.45\linewidth]{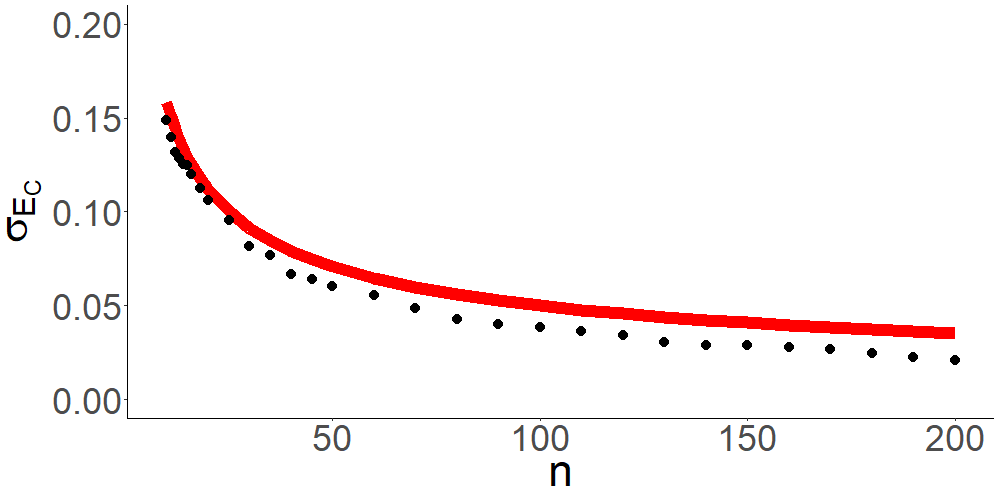}~\label{sigma_E_C_n_CR}}
 \subfloat[]{\includegraphics[width=0.45\linewidth]{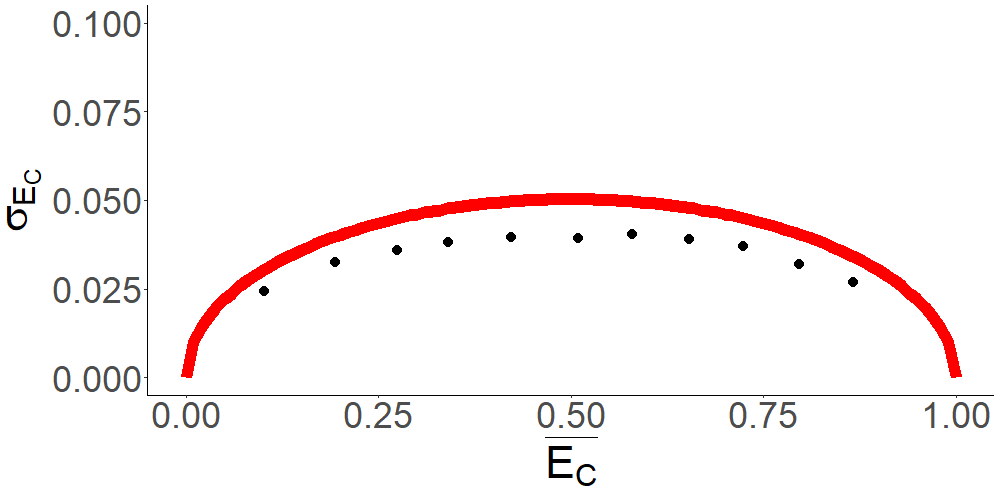}~\label{sigma_E_C_p_CR}}
  \caption{Comparison between empirical data (black points) and Eqn. \ref{eqn:sigma_EC_stats} (red lines). a) Standard deviation of $E_C$ as $n$, the number of responses in each sample test set, varies empirically (code: CR, $p_{train} = 0.5$, $\overline{E_C}\approx 0.5$, and the proportion of responses in the test bank that contain the code is 0.5).  b) Standard deviation of $E_C$ as $\overline{E_C}$ varies empirically (code: CR, $n = 100$, $p_{train} = 0.5$).~\label{sigma_E_C_CR}}\vspace{-1em}
\end{figure*}

\subsection{Explanation of method}

To quantify the variation in the measurements made by machine coding, we calculate $\sigma_{E_C}$, the standard deviation of a distribution of values of $E_C$ drawn from many different sample test sets of size $n$ pulled from a larger test bank of size $N_{bank}$. 
We expect some fraction of the measurements of $E_C$ to fall within one standard deviation of the mean $\overline{E_C}$ (approximately 68\% if the distribution is normally distributed). We, therefore, define the uncertainty in a single measurement of $E_C$ as the standard deviation of this distribution, as it captures the typical variability between individual measurements. While we can produce this estimate from repeated tests of a large data set, we need a way for researchers to estimate the value of $\sigma_{E_C}$ from a single data set (such as when applying a previously trained algorithm to new data without human-coding).

Fortunately, this estimation is possible because the standard deviation, $\sigma_{E_C}$, is characteristic of the phenomenon we are measuring. Because we are using an OVA approach, such that each response either contains ($1$) or does not contain ($0$) the code, our data are dichotomous. Thus, $E_C$, the computer's estimate of the frequency the code appears in a particular sample test set, can be represented as the mean of a Bernoulli distribution, where the number of times the code is present in the sample is $n_p = nE_C$ and the number of times the code is absent in the sample is $n_a = n(1-E_C)$. The standard deviation $\sigma_B$ of the Bernoulli distribution can be calculated as: 

\begin{equation}
    \sigma_B^2 = \frac{1}{n} [n_a(0-E_C)^2 + n_p(1-E_C)^2] = E_C (1-E_C).
\end{equation}

The central limit theorem implies that $E_C$ comes from a distribution with mean $\overline{E_C}$ and standard deviation $\overline{\sigma}_B/\sqrt{n}$, where $\overline{\sigma}_B$ is the average value of $\sigma_B$ that emerges across all sample test sets. Thus, the statistical uncertainty in a single measurement of $E_C$ is equal to $\overline{\sigma}_B/\sqrt{n}$. When only a single measurement of $E_C$ has been taken (such as with new data not coded by humans) and $\overline{E_C}$ is unknown, we can approximate that $\overline{E_C}\approx E_C$ and estimate $\sigma_{E_C}$ directly: 

\begin{equation}
\sigma_{E_C} = \overline{\sigma}_B/\sqrt{n} = \sqrt{\frac{\overline{E_C}(1-\overline{E_C})}{n}} \approx \sqrt{\frac{E_C(1-E_C)}{n}}.
\label{eqn:sigma_EC_stats}
\end{equation}

This estimate can be computed with a single sample test set, as it requires only the measurement $E_C$ in a single sample test set and the number of responses $n$ in that sample test set. Importantly, this statistical uncertainty can be applied to a machine-coded sample test set \emph{without any new human-coding} beyond that used to initially train the model. 
Below, we evaluate this estimate of statistical uncertainty with empirical results to demonstrate its robustness to finite sampling.  

\subsection{Evidence and Examples}

We assess the validity of Eqn. \ref{eqn:sigma_EC_stats} in two steps. First, we evaluate the relationship between $\sigma_{E_C}$ and $n$. Second, we evaluate the relationship between $\sigma_{E_C}$ and ${\overline{E_C}}$. 

To evaluate the relationship between $\sigma_{E_C}$ and $n$ (sample test set size), we train an algorithm for the CR code using a training set of size $600$ responses with $p_{train} = 0.5$ (as per the recommendation to balance the training set, described above in the methods section). We test 30 different values of $n$ between 10 and 200. For each value of $n$, we draw 1000 sample test sets of size $n$ from a test bank where the proportion of responses containing the code is also 0.5 (so that we are controlling for variability due to proportion of responses in the test set sample). We then have our algorithm compute $E_C$ for every sample. For each value of $n$, we calculate an empirical value for $\sigma_{E_C}$ by computing the standard deviation across the 1000 samples. 

We find that the empirical values of $\sigma_{E_C}$ approximately follow the predicted $1/\sqrt{n}$ relationship (see Fig. \ref{sigma_E_C_n_CR}), though the empirical values are consistently less than the values predicted by Eqn. \ref{eqn:sigma_EC_stats}. This overestimation may be because our test set samples are pulled from a test bank of finite size ($N_{bank} = 300$), so the variability does not quite reach the level we would expect if our samples were pulled from a large population. That is, the number of ways data may vary in different samples is diminished to some extent by the fact that the test bank is only a few times larger than $n$. Fortunately, Eqn. \ref{eqn:sigma_EC_stats} over-predicts, rather than under-predicts, the standard deviation, meaning the calculated $\sigma_{E_C}$ from a single test set sample will be a conservative estimate of statistical variability. 

To evaluate the relationship between $\sigma_{E_C}$ and $\overline{E_C}$, we test values of $\overline{E_C}$ between 0 and 1. We use the same algorithm for the CR code with a training set of size $600$ and $p_{train} = 0.5$. We create a set of test banks where the proportion of responses containing the code (as determined by the human coder) ranges from 0 to 1 inclusive in steps of 0.1. We draw repeated samples from each test bank; each of these values corresponds to a fixed value of $\overline{E_C}$ (these values range from about 0.1 to 0.9 as systematics narrow the range). For each value of $\overline{E_C}$, we take 1000 sample test sets of size $n = 100$ and calculate $\sigma_{E_C}$ across the 1000 samples.

We find that the empirical values of $\sigma_{E_C}$ again approximately follow the predicted parabolic relationship (see Fig. \ref{sigma_E_C_p_CR}). As seen in the relationship to $n$, the empirical values are consistently less than the values predicted by Eqn. \ref{eqn:sigma_EC_stats}, likely for the same reason. Again, the analysis shows that Eqn. \ref{eqn:sigma_EC_stats} provides a conservative estimate of statistical variability.

In our supplemental materials, we demonstrate the validity of estimating $\sigma_{E_C}$ through a measurement of $E_C$ from a single sample test set rather than the mean $\overline{E_C}$. We also show, for multiple coding schemes and values of $n$, that this statistical variability is independent of the size of the training set (Fig. \ref{sigma_E_C_vs_train_size}). This independence of variability with training set size reflects that the statistical variability comes only from sampling a subset of responses from the larger test set. This means that $\sigma_{E_C}$ is constant even if the size of the training set is so small that past work would suggest the machine coding would be quite poor~\cite{Fussell_etal}.

Finally, we provide an example of calculating statistical uncertainty using this method. Consider a dataset of 160 responses that have not been coded by humans. Without performing any human coding, we can apply our trained model to the 160 responses. Say the trained model applies the code to 96 of the 160 responses. This means that $E_C = 96/160 = 0.6$. We take this value of $E_C$ as an approximation of $\overline{E_C}$. Therefore, $\sigma_{E_C} \approx \sqrt{0.6*0.4/160} = 0.039$.

In summary, $\sigma_{E_C}$ can be conservatively estimated from two characteristics of a sample test set: the computer's estimate of the presence of the code in the sample test set, $E_C$, and the number of responses in the sample test set, $n$. The researcher can then report their estimate of the presence of the code as $E_C \pm \sigma_{E_C}$. 

\subsection{Recommendations}
The analysis above leads to a single recommendation for estimating statistical uncertainty:

\begin{enumerate}
    \item Whenever reporting the frequency of a code in a sample estimated by machine coding, report the statistical uncertainty using Eqn. \ref{eqn:sigma_EC_stats}.
\end{enumerate}

\noindent We note that it is possible to do this without any human-coding beyond the initially human-coded training set. 



\section{Systematic Uncertainty from the trained algorithm}\label{Sec:systematic}

The third step in establishing the trustworthiness of an algorithm is to understand any potential systematic effects in how the algorithm codes data. When comparing two coders, such as a machine coder and a human coder or even two human coders, the two coders can apply the same coding scheme in consistently different ways, leading to systematic (as opposed to random) uncertainty. In this section, we demonstrate how systematic effects, wherein a machine coder can systematically over- or underestimate the presence or absence of a code, emerge due to an outcome imbalance in the sample test set ($E_C$). We then show how this systematic effect can be measured and thus used to calibrate a measurement from a trained algorithm. For now, we consider only systematics when applying an algorithm to data from a similar population. The next section will explore impacts due to data from different populations.


\subsection{Explanation of method}

We can evaluate potential systematic offsets between the human and machine coder by calculating the difference between an individual measurement of $E_C$ and the human estimate $E_H$. We observe that the computer systematically over- or under-estimates the presence of a code depending on the prevalence of the code in the test bank.
We can estimate this relationship between $E_C$ and the systematic $E_C - E_H$ through a predictive linear model, which we define as a function $S(E_C)$.
Following the determination of this best fit function $S(E_C)$ from a set of human-coded test banks, researchers can then apply this function to measurements without further human-coding. We assume that the human coder is ``correct'' whenever the human and machine coders disagree, such that we calibrate the machine coder to match the human coder (rather than the other way around). 


\subsection{Evidence and Examples}

\begin{figure*}
  \subfloat[]{\includegraphics[width=0.45\linewidth]{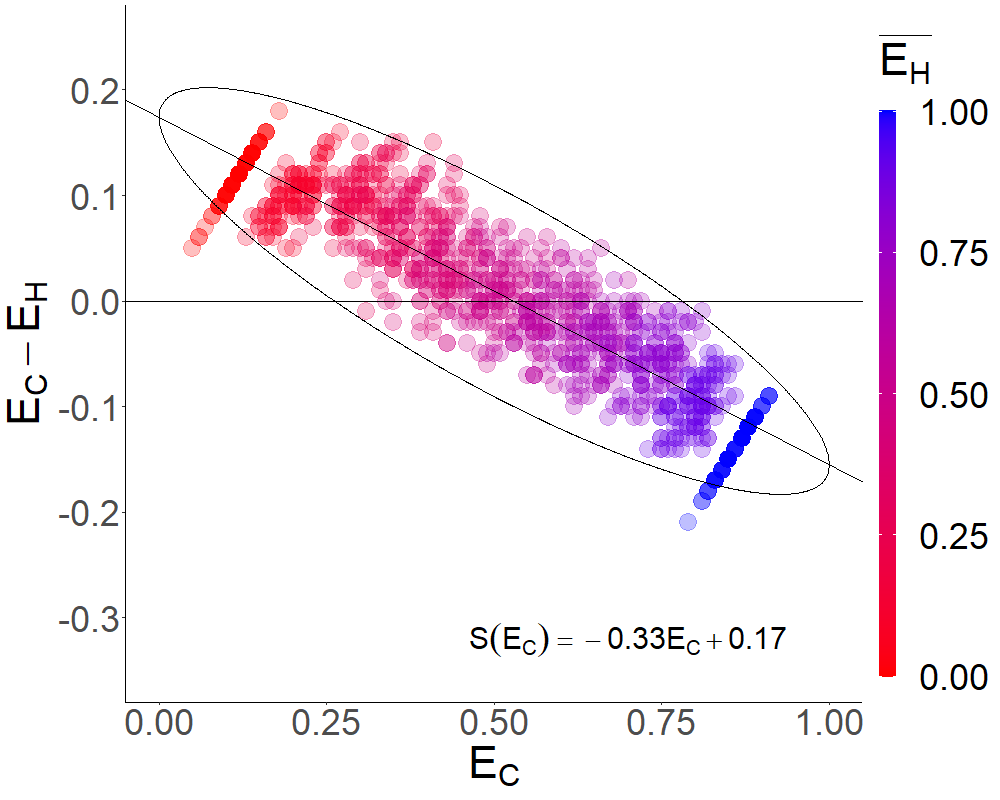}\label{systematics_CR}}
 \subfloat[]{\includegraphics[width=0.45\linewidth]{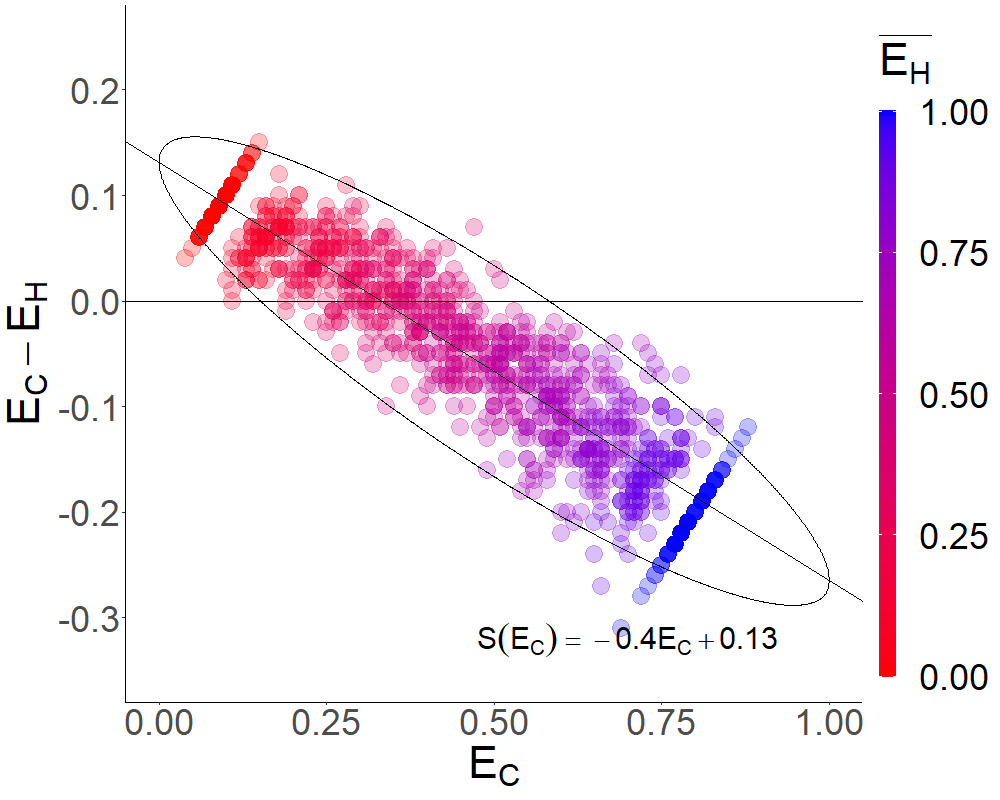}\label{systematics_L}}
  \caption{Given a value of $E_C$, we compute the systematic using the line of best fit $S(E_C)$. Scatter around the line of best fit is within the statistical uncertainty, with at least 95\% of the data falling within the black oval depicting $2*\sigma_{E_C}$ (see Eqn.\ref{eqn:sigma_EC_stats}. a) $S(E_C)$ for the CR code b) $S(E_C)$ for the L code.~\label{systematics}}\vspace{-1em}
\end{figure*}


We calculate the systematic effect with trained models for both the CR and L codes. Following the recommendation above, we fix $p_{train} = 0.5$, which has the additional benefit of controlling for a relationship between $p_{train}$ and the systematic effect (as observed in Sec. \ref{optimizing}). 
We generate a set of eleven different test banks where the proportion of responses containing the code varies from $0$ to $1$ inclusive in steps of $0.1$. The size of each test bank is $N_{bank} = 200$. For each test bank, we compute $E_C$ and $E_H$ for 100 sample test sets of size $n = 100$ pulled from the test bank. 

We observe a systematic difference between the human and machine coder that tends to be more extreme with more extreme values of $E_C$ (Fig. \ref{systematics}). For the CR code (Fig.~\ref{systematics_CR}), the systematic is such that the model overestimates the prevalence of the code when $E_C < 0.5$ and underestimates the prevalence of the code when $E_C > 0.5$. For the L code (Fig.~\ref{systematics_L}), the direction of the systematic is the same, but the boundary between over- and underestimation is shifted: the model overestimates the prevalence of the code when $E_C < 0.3$ and underestimates the prevalence of the code when $E_C > 0.3$. Variability around the average systematic difference falls within the previously defined statistical uncertainty, as 95\% of the data falls within two standard deviations of $\sigma_{E_C}$ as calculated by Eqn. \ref{eqn:sigma_EC_stats} (indicated by the black ovals in Fig. \ref{systematics}). These results imply that a measurement of $E_C$ is subject to a systematic error that itself is dependent on the value of $E_C$.

The effect is also dependent on the trained model, which we demonstrate in our supplemental materials (Fig. \ref{systematics_other_training}). There, we see that when we construct new trained models for the CR and L codes using a different (yet still balanced) training set, the best fit lines, $S(E_C)$, have the same overall downward linear shape but have different parameters than the fit equations in Fig. \ref{systematics}. The estimates for the systematic, therefore, depends on the training set used (i.e., is unique to the trained model). 
The analysis also shows that the difference between systematics for two different trained models can go beyond the statistical uncertainty, especially at extremes. We expect that this occurs because each trained model learns slightly different coding rules. 



To account for this source of systematic uncertainty in subsequent analysis, a researcher can use the best fit function for their trained model to find the systematic offset that corresponds to their measured value of $E_C$ for a new data set (that has not been coded by humans). One should report the estimate of $S(E_C)$ for their estimate of $E_C$ along with its corresponding statistical uncertainty as $E_C \pm \sigma_{E_C} - S(E_C)$. For example, for a measurement of $E_C = 0.75$ with the CR code (and the trained algorithm that produced Fig.~\ref{systematics_CR}), we would use the equation for $S(E_C)$ in Fig.~\ref{systematics} to obtain $S(E_C=0.75) = -0.33\times0.75 + 0.17 = -0.08$. 



\subsection{Recommendations}

We identify two key recommendations for determining the systematic effects in machine coding:
\begin{enumerate}
\item{For a single trained model applied to new data from a similar population, measure the best fit equation $S(E_C)$ for the systematic uncertainty as a function of the machine coding estimate, $E_C$, using a test bank of human-coded data. 
The larger the test bank size and sample test set size and the larger the spread in values of $E_C$, the more confident you will be in the function.}
\item{
Report the systematic uncertainty alongside your estimate and its statistical uncertainty in the form $E_C \pm \sigma_{E_C} - S(E_C)$.}

\end{enumerate}
\noindent If the magnitude of the systematic uncertainty of your result is large enough that it calls your results into question, improve your trained model with more data or through improvements to the natural language processing and machine learning.

We return to the question of how much hand-coded data are needed to measure systematics in this way. In our analysis of the CR and L codes, we used test data that included at least 200 responses that contained the code and at least 200 responses that did not (separate from the 600 responses included in the training set). This full amount of 400 responses may or may not be necessary for researchers to evaluate systematics, depending on the particular code. One way this number could be reduced is by reducing the range of $\overline{E_H}$ values in the test banks, though there is a trade-off as this will reduce precision in the calculated systematic value. In Fig.~\ref{systematics} we use the full available range of values, as we include a test bank with $\overline{E_H} = 0$ and a test bank with $\overline{E_H} = 1$. If the prevalence of a code ranges from, say 0.2 to 0.6, and depending on educational conditions, it may not be necessary to calculate the best fit function $S(E_C)$ on this full range. Alternatively, one can reduce the size of the test banks at each value of $\overline{E_H}$, which also reduces precision but would maintain the full range.

\section{Systematic uncertainty in new data sets}\label{Sec:systematic_new}

The fourth step in establishing the trustworthiness of an algorithm is to evaluate any potential systematic effects based on how an algorithm may code data that comes from a distinct population of, in our case, student responses than those used in the initial training. Quantitatively estimating this source of potential systematic offsets is particularly important for physics education researchers looking to use machine learning and natural language processing at scale, such that a model trained on data collected at a particular time, institution, course, or course level could then be applied to large quantities of data from other contexts with limited additional human coding. In this section, we demonstrate how to estimate such systematics using similar methods to those in the previous section.

\begin{figure*}
  \subfloat[]{\includegraphics[width=0.45\linewidth]{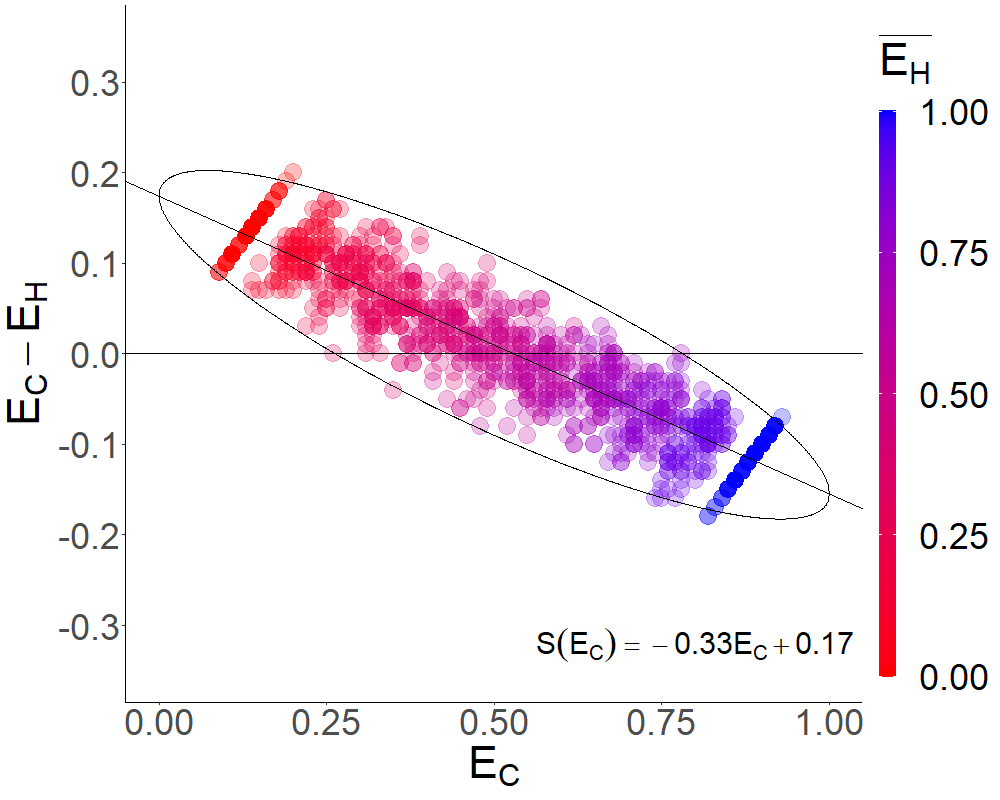}}
 \subfloat[]{\includegraphics[width=0.45\linewidth]{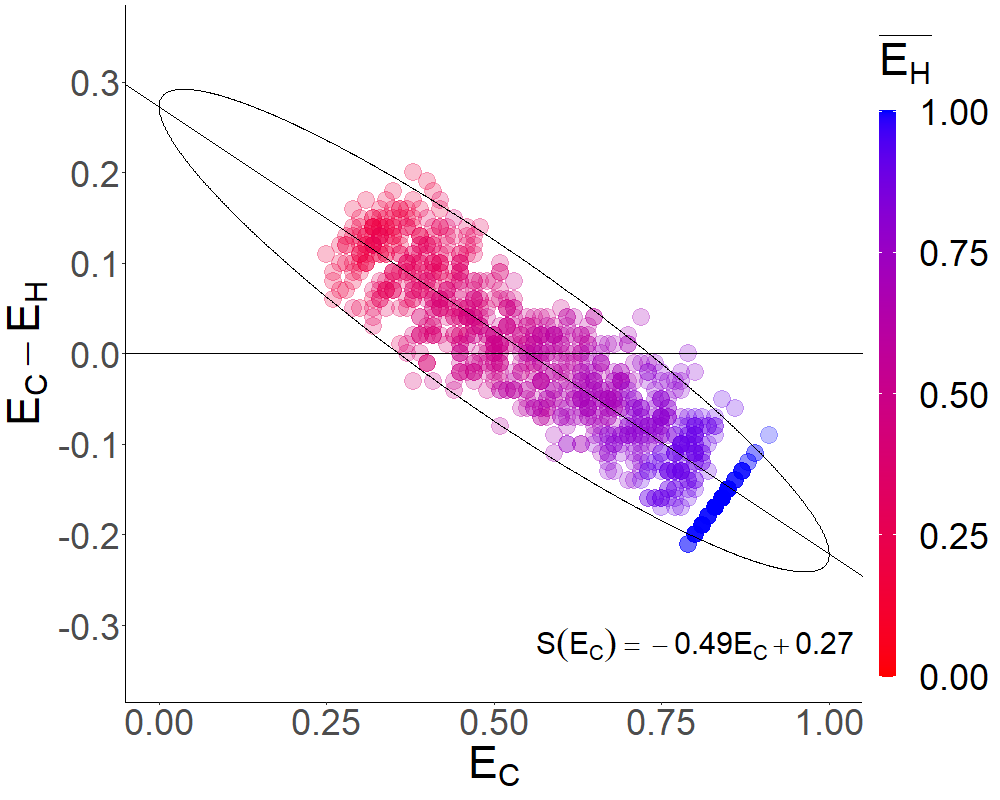}}
  \caption{Sample test sets from the same population as the training set have lower systematics than sample test sets from a different population as the training set. The training set is identical in both plots, consisting of a 20/80 mix of pre/post responses. a) $S(E_C)$ with sample test sets from same population as the training set (test bank is 20/80 mix of pre/post responses). b) $S(E_C)$ with sample test sets from different population as the training set (test bank is all pre responses). (Code = CR, $p_{train} = 0.5$).~\label{population_systematics}}\vspace{-1em}
\end{figure*}

\subsection{Explanation of method}

In the previous section, we demonstrated that systematic effects between the human and machine coder can be written as a function of $E_C$ when the training set and test set are pulled from the same population. Additional systematic effects may arise, however, when the characteristics of students in the test set are different from those represented in the training set \cite{Young_Caballero}. These characteristics can impact the machine coding if, for example, the new population of students use language differently than the initial data set. For example, students may use unique jargon or special terms that a particular instructor introduced in a particular physics class, but not in others. There are, of course, many other ways in which the responses by one population of students may differ systematically from those of another population.

To account for this additional systematic effect, we determine the function $S(E_C)$ using human-coded test data from the new population. When reporting results, accounting for the systematic then follows the steps from the previous section. Though some new human-coding is needed to complete this step, we emphasize that this method limits human-coding to only the amount needed to establish trust in the researchers' ability to measure systematic effects in the new data set. Next, we will provide an example of estimating the function $S(E_C)$ with data from a new population and we will compare it to an estimate of $S(E_C)$ with data from the same population as the training set. 

\subsection{Evidence and Examples}

Our Trustworthy data set includes several semesters where the Trustworthy question was asked only at post-survey. Because the survey question is related to course material for the populations being surveyed, we expect that there are meaningfully different characteristics between the pre-survey and post-survey responses. 
This provides an opportunity to test for population systematics by assuming that responses from students at pre-survey represent a distinct population than responses from students at post-survey. To do so, we generate a trained model for the CR code with a training set size of 500 majority post-survey responses (in this instance we had to reduce training set size and could not use exclusively post-survey responses because of limited data). The training set is comprised of 80\% post-survey responses and 20\% pre-survey responses. In accordance with the recommendations above, we balance the training set such that half the responses contained the CR code according to human coders. 

Using non-training data responses remaining in the data set, we create two sets of test data: the first set  (Set Pre/Post) mirrors the structure of the training set with 80\% post-survey and 20\% pre-survey responses, while the second set (Set Pre) includes only pre-survey responses. Thus, Set Pre/Post is considered the same population as the training set and Set Pre is considered a different population. 

We calculate the fit $S(E_C)$ as a function of the computer's estimate $E_C$ using the methods outlined in the previous section. We construct test banks where the proportion of responses that contain the code varies from 0 to 1. Each test bank is of size $N_{bank} = 200$. From each test bank, we take sample test sets of size $n = 100$. We could not generate samples with values of $E_C$ lower than about $0.3$ using Set Pre because of insufficient pre-survey data. 

As per the previous section, we again find a linear trend between the systematic $E_C - E_H$ and the computer's estimate $E_C$. We find the systematics for Set Post/Pre are given by the fit line $S(E_C) = -0.329*E_C + 0.175$ (Fig. \ref{population_systematics} a) and the systematics for Set Pre are given by the fit line $S(E_C) = -0.494*E_C + 0.273$ (Fig. \ref{population_systematics} b). 
Across the range of possible values of $E_C$, the systematic tends to be farther from zero when machine coding samples from the Set Pre, a different population than the training set, compared to when machine coding samples from the Set Post/Pre, the same population as the training set. We also see that the statistical variability is the same for both Set Pre and Set Pre/Post and that 95\% of the data fall within two standard deviations as calculated by Eqn.~\ref{eqn:sigma_EC_stats} (given by the black ovals in Fig. \ref{population_systematics}).

We have presented just one example of the systematic differences that may arise when the characteristics of students in the test set are different from those represented in the training set. The results motivate caution to be taken in all instances involving changes to population characteristics between training and test data, including changes in the composition of, for example, student major, institution, gender, race and ethnicity, or international student status. Many variables can affect the way language is used and thus how machine (and human) coders interpret it. We have shown, however, that systematic differences can be measured by hand-coding a relatively small set of data from the new population. 
Future work should continue to investigate the effects that arise from applying machine coding to different populations of students. 

\subsection{Recommendations}
We identify one key recommendation for determining the systematic effects in applying a machine coding algorithm to a new population:
\begin{enumerate}
\item When applying a trained model to a new student population, re-generate $S(E_C)$ by hand-coding a subset of the test data from the new population. Then follow the recommendations of the previous section using this new function $S(E_C)$. 
\end{enumerate}

Additional hand-coded data are needed to measure systematics in the new data set. The amount of data needed will be comparable to the amount of data needed to measure systematic uncertainty from the trained algorithm in Sec.~\ref{Sec:systematic}, though it may be possible to use less data if it can be shown with a smaller amount of data that systematics are not greater than the previously calculated systematic uncertainty in the trained algorithm. In this case, the previously calculated function $S(E_C)$ may be used to compute the systematic.



\section{Fully worked example}\label{Sec:worked_example}
\begin{figure}{\includegraphics[width=0.9\linewidth]{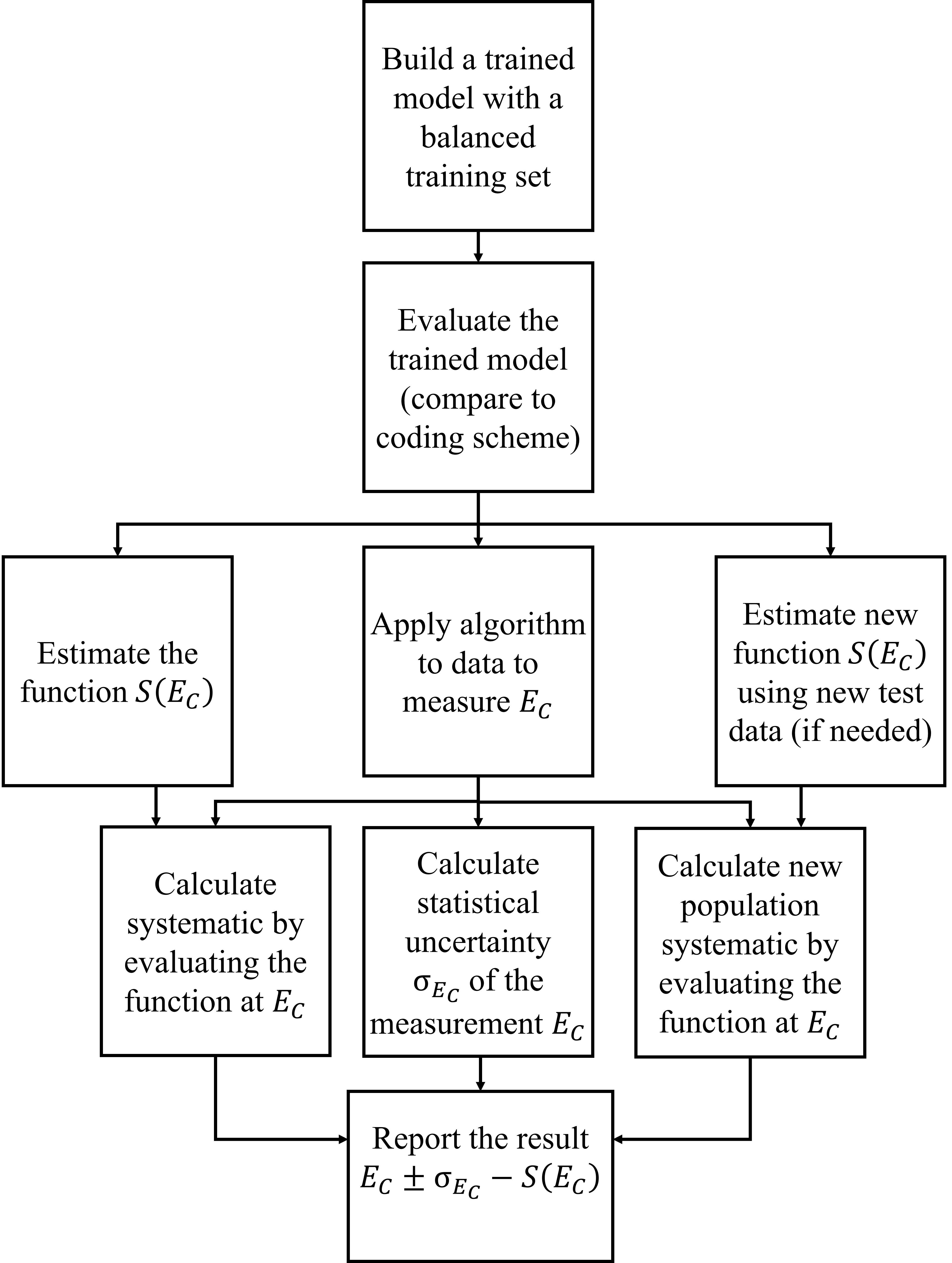}}
  \caption{Flowchart depicting the stages of the methodology.~\label{flowchart}}\vspace{-1em}
\end{figure}
As a demonstration of this methodology altogether, this section presents a fully worked example with a different code than in the previous analyses. In what follows, we train a new model, evaluate the optimized model, and calculate the statistical and systematic uncertainties associated with an estimate of the code in a new test set without additional human coding, following the methodology summarized in Fig.~\ref{flowchart}. 

For this example, we use a different code from the Trustworthy question coding scheme: Uncertainty, abbreviated U. Two human coders achieved a Cohen's kappa value of 0.9 in 10\% of the data for this code. Then one coder coded the rest of the data. The U code is defined as: ``measures were taken during the procedure to reduce/account for error/uncertainty and/or there is a small calculated uncertainty.'' The inclusion criteria for the U code are: 
\begin{enumerate}
    \item uncertainty in measurements, 
    \item uncertainty of the results, 
    \item methods serve to reduce error, 
    \item error bars/bounds, 
    \item sources of error in measurements, 
    \item signal-to-noise ratio, 
    \item accounting for or reducing systematic error, random error, or ``human error'' in the process of taking measurements. 
\end{enumerate}
We use four distinct data groupings for this analysis: a training set ($400$ pre- and post-survey responses from semesters prior to Fall 2022), 
a new data set not coded by humans ($286$ post-survey responses from Fall 2022), and two sets of test data. The first set of test data (labelled Set Pre/Post) includes responses from a similar population to those in the training set data but different from those in the new (un-coded) data ($270$ pre- and post-survey responses from semesters prior to Fall 2022). The second set of test data (labelled Set Post) includes responses from a different population to those in the training data but similar to those in the new (un-coded) data ($270$ post-survey responses from semesters prior to Fall 2022). The responses for the training data and the test data were all hand-coded but the new (post-survey from Fall 2022) data set was not.

We first trained an algorithm for the U code using a training set of size $400$. We set $p_{train} = 0.5$ as per the recommendations (in this instance we had to reduce training set size because of limited data, as instances of the U code were less common).

\subsection{Evaluate the trained model}

 \begin{figure*}
  \subfloat[]{\includegraphics[width=0.45\linewidth]{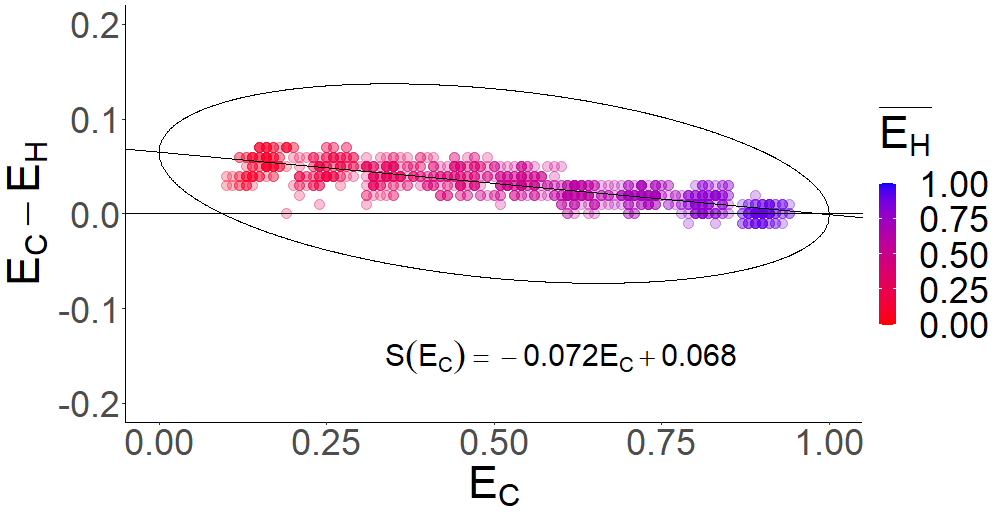}}
 \subfloat[]{\includegraphics[width=0.45\linewidth]{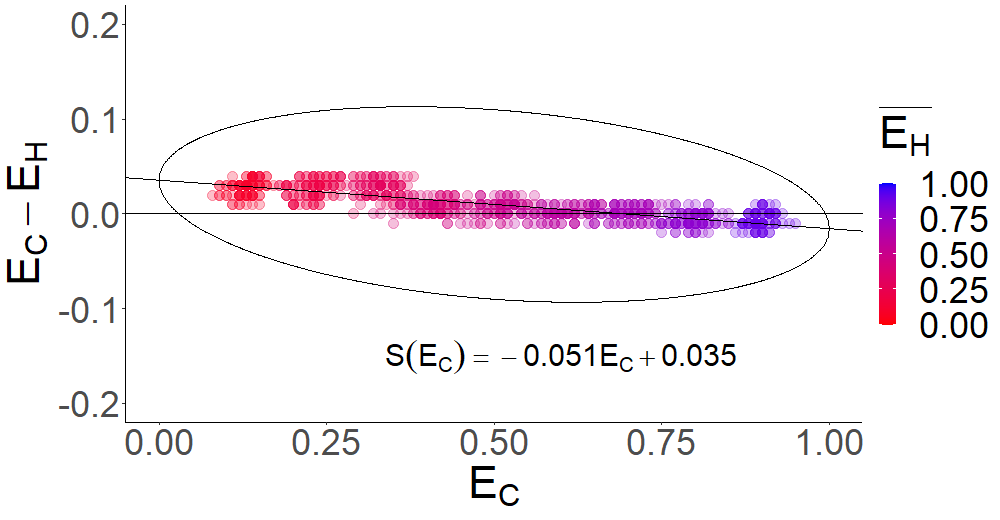}}
  \caption{Trained model for the U code applied to sample test sets from two different sets of test banks: a) $S(E_C)$ with Set Post/Pre, and b) $S(E_C)$ with Set Post (Code = U, $p_{train} = 0.5$. Black oval depicts 2 * $\sigma_{E_C})$.~\label{uncertainty_code_systematics}}\vspace{-1em}
\end{figure*}
Table~\ref{tab:uncertainty} presents the coefficients associated with the trained model. The highest weighted words are ``uncertainty'' and ``error'', which are part of six of the seven inclusion criteria for this code, such as accounting for or reducing error, human error, and sources of error. One inclusion criterion, ``signal to noise ratio,'' occurs rarely in the data set and does not appear in the top weighted words. Though the U code does not have explicit exclusion criteria, ``percent error'' was part of the inclusion criteria for a different code and was understood by the human coders to not be related to the inclusion criteria for the U code. Accordingly, ``percent'' is the one of the words with the lowest (most negative) coefficients. Overall, the trained model seems to accurately capture our coding scheme based on these coefficients. 

\begin{table}[!htbp]
  \caption{Coefficients for the trained model for the U code. Left side (coefficient > 0) corresponds to inclusion criteria and right side (coefficient < 0) corresponds to exclusion criteria. Words in bold are listed in the inclusion criteria for the human-generated coding scheme for this code (or share a root and are part of the same word family as a word in the inclusion criteria, e.g. ``uncertainty'' and the common typo ``uncertainities'').\label{tab:uncertainty}}
  \begin{ruledtabular}
    \begin{tabularx}{\textwidth}{c c | c c }
      \textbf{Word} & \textbf{Coefficient} & \textbf{Word} & \textbf{Coefficient} \\ 
      \hline
      \textbf{uncertainty} & 4.76 & percent & -0.97\\
      \textbf{error} & 3.09 & this & -0.58 \\
      low & 1.07 & within & -0.56 \\
      \textbf{human} & 0.92 & fit & -0.50\\
      \textbf{source} & 0.84 & it & -0.49\\
      \textbf{uncertainities} & 0.70 & likely & -0.47\\
      \textbf{account} & 0.68 & consistent & -0.47\\
      minimized & 0.60 & theoretical & -0.46\\
      possible & 0.58 & analysis & -0.44\\
      \textbf{accounted} & 0.55 & peer & -0.42\\
      \textbf{reduce} & 0.53 & same & -0.39\\
    \end{tabularx}
  \end{ruledtabular}\vspace{-1em}
\end{table}

\subsection{Calculate statistical uncertainty}

We apply the trained model to $n = 286$ responses in the un-coded data set. This machine coding returns a measurement of $E_C = 0.409$. With this measurement, we calculate the statistical uncertainty from Eqn. \ref{eqn:sigma_EC_stats}, giving $\sigma_{E_C} = \sqrt{0.409*(1-0.409)/286} = 0.029$. 

\subsection{Calculate systematic uncertainties}

To calculate systematic uncertainty associated with the new un-coded data, we first need to determine a systematic uncertainty function from the hand-coded test data. First, we divide Set Pre/Post (where data come from a population similar to the training data but different from the un-coded data) into nine different test banks, each with a different proportion of responses containing the code (according to human coders) between $0.1$ and $0.9$ inclusive. The size of each test bank was $N_{bank} = 150$. For each test bank, we compute $E_C$ and $E_H$ for 100 sample test sets of size $n = 100$ pulled from the bank. Again, because U is a less common code, we could not make larger banks without losing the ability to estimate $S(E_C)$ on as wide a range between $E_C$ = 0 and 1 as possible. 

We generate a plot of $E_C - E_H$ versus $E_C$  (Fig. \ref{uncertainty_code_systematics} a) for the Set Pre/Post data. The systematics can be measured with the fit line $S(E_C) =-0.072*E_C + 0.068$(Fig. \ref{uncertainty_code_systematics} a). The systematics follow the same general pattern (negative slope) seen previously, but the range is smaller than for codes we discussed previously. The observed statistical variability of $E_C - E_H$ is also much smaller than the statistical uncertainty of $E_C$, with nearly all data falling well within two standard deviations (given by the black oval). We expect that this reduced variability in $E_C - E_H$ reflects that the code is particularly and confidently captured by the trained model.\footnote{The ovals in Fig.~\ref{uncertainty_code_systematics} represent the statistical uncertainty of $E_C$, while the variability in the data about the line of best fit reflects the statistical uncertainty of the quantity $E_{C} - E_{H}$. The variability of $E_C$ - $E_H$ should be less than or equal to our prediction for the statistical uncertainty of $E_C$ because $E_C$ and $E_H$ are not independent variables (they are correlated). 
The purpose of adding the ovals is not to provide a theoretical prediction for the variability in $E_{C} - E_{H}$, but rather to provide a check that the variability in the systematic of the measurement $E_{C}$ does not exceed what is already being reported as the statistical uncertainty for the measurement $E_{C}$.}  

We then perform the same analysis using Set Post, where data come from a population different from the training data but similar to the un-coded data. In this case, the systematic can be measured with the fit line $S(E_C) = -0.051*E_C + 0.0353$ (Fig. \ref{uncertainty_code_systematics} b). We again see the same linear trend, but now with a slightly smaller slope magnitude and a smaller intercept. 
This analysis demonstrates that, in this example, perhaps counterintuitively, the systematic effect is smaller when estimated using hand-coded test data from a population \textit{different} from the training data, though this difference is indistinguishable given the statistical uncertainty.  

After doing this preparatory work, we calculate the systematic uncertainty of the trained model with our estimate of $E_C = 0.409$. The systematic uncertainty from the Set Pre/Post analysis is $S(E_C = 0.409) =-0.072*0.409 + 0.068 = 0.039$.
Next, we calculate the systematic uncertainty using Set Post because our un-coded data are post-survey data. The systematic uncertainty from the Set Post analysis is $S(E_C = 0.409) =-0.051*0.409 + 0.0353 = 0.014$. 


\subsection{Reporting the result}
Based on our methodology, we would report that the frequency of the U code in the new data set is $E_C = 0.409 - 0.014 \pm 0.029$. We opt to use the estimate of systematic uncertainty from Set Post rather than Set Pre/Post because the student population in Set Post is more similar to the population in the new data set. The difference between the two calculations of systematic uncertainty are indistinguishable within the larger statistical uncertainty, so we consider the estimate of systematic uncertainty from Set Post to also capture any systematic uncertainty from the trained algorithm.

The skeptical reader may ask if this measurement aligns with one made through human coding of the same dataset. We do not wish to promote comparison with human coding as a validation method because it eliminates our ability to use machine coding as a tool to improve efficiency. We can report, however, that upon skimming through the spreadsheet of the machine's codes, the machine coding of individual responses appeared correct. Nonetheless, to appease the skeptical reader, we human coded the first 60 responses in the un-coded new data set and measured $E_H =  0.419 \pm 0.064$, where we calculate statistical uncertainty using Eqn. \ref{eqn:sigma_EC_stats}. This result is well within uncertainties (both statistical and systematic) of the computer estimate. 

It is notable that the systematic uncertainty for the U code (Fig.~\ref{uncertainty_code_systematics}) is significantly reduced compared to the systematic uncertainty for the codes we examined earlier (Fig.~\ref{systematics}). Furthermore, the level of actual statistical variation about the systematic fit is smaller than in the systematic fits for the other codes (Fig.~\ref{systematics}). We interpret this reduced uncertainty to be a property of the U code, specifically. That is, the trained model was able to better capture the U code, likely because sentences that contain this code tend to be more similar to one another compared to the CR and L codes.

\section{Future Work}\label{Sec:future_work}

We have presented a framework for measuring and reporting quantitative education claims with machine coding of student text data. This framework was developed using two example data sets, the Trustworthy data set and the Sources data set, and only a handful of individual codes. The limitations of human-coding constrained the number of data sets and codes we could include in this analysis. As this framework continues to be applied, new data sources and coding schemes should be used to reveal nuances that improve and build upon our recommendations. 

There are specific use cases of supervised natural language processing where these methods cannot be used. Most notably, these methods cannot be used to assist in making inferences about individual students, because these methods are designed to estimate the prevalence of a code on a population level and to compute statistical and systematic uncertainty of this estimate. Rather, these methods should be used to evaluate samples of responses from many students, such as individual classes.

The analyses above also relied on a one-vs-all approach, where each code was evaluated independently and a response could be coded for any number of codes. It is possible to use a one-vs-all approach with a coding scheme that uses multiple mutually exclusive codes, as in Ref.~\cite{Wilson_etal_2022}. There may be some cases, however, where it is necessary to enforce the constraint that the prevalence of these mutually exclusive codes must add to one. Future work should evaluate and adapt our methods for such cases. The estimate of statistical uncertainty, for example, would be different because the statistical processes would come from a different probability distribution than the Bernoulli distribution. 

Our machine coding algorithms also used logistic regression and our method for evaluating the correspondence between the trained model and human-generated coding scheme in particular is based in the logistic regression algorithm. Future work may use and expand upon these methods by using other algorithms, such as RandomForest, SVM, and neural networks. Additionally, future work should investigate empirically if statistical and systematic uncertainty are impacted by the use of these other algorithms. One particularly promising avenue, as pre-trained Large Language Models (LLMs) become available, is to integrate LLMs into the construction of trained models that classify PER data. Evaluating the correspondence between trained models that use LLMs and a human-generated coding scheme may pose a challenge because you cannot read off parameters associated with unique features as you can for logistic regression. This may be worthwhile, however, as the LLM approach could reduce the amount of training data needed while also reducing systematic uncertainty. With the LLM approach, it is still necessary to account for statistical and systematic uncertainty in analysis and we expect the methods above would apply. 

A key constraint on the methods throughout is the amount of hand-coded data required. An untested idea is that generative AI could be used to generate additional responses that do or do not contain a code (in a process similar to efficient mass creation of isomorphic physics problems~\cite{chen2023reforming}). These additional responses could particularly be used to create much larger balanced training sets and larger test sets for assessing systematic effects. The methodological and ethical considerations that would go into this process are beyond the scope of this paper, but we would be interested in seeing this idea explored in future work.

We focused our analysis of systematic effects from the trained algorithm on the role of outcome imbalance in the test set. Other sources of systematic uncertainty that we did not explore may also exist, which is another area of future study. In addition, we demonstrated how a difference in student characteristics (whether they are taking a pre survey or a post survey) impacts the level of systematics. Future work should investigate explicitly the extent to which other characteristics (gender, race, institution, course style) measurably impact systematic uncertainty for a range of data sources. For example, AI detectors have been shown to systematically label writing from non-native English speakers incorrectly as AI-generated~\cite{Liang_etal}, so similar systematic errors are likely to occur when using machine coding in PER.


Lastly, there may be ways to reduce the amount of additional hand-coded data needed to compute systematic error while making trustworthy claims. For example, a researcher may be able to dramatically reduce the amount of hand-coded data needed for evaluating systematics if they are willing to accept the cost of increased systematic error attributed to their claims. A researcher may choose not to find the best fit function $S(E_C)$ for a particular code and student population group if they can instead provide evidence to suggest that the systematic uncertainty is well within the statistical variability from random sampling. For example, it could be argued using approximately 50 responses that the U code has a systematic uncertainty much less than the statistical uncertainty. As with any research that seeks to make claims about large groups of students, there is a fundamental tension between the additional effort needed to gather additional data and the fact that gathering additional data will usually improve the trustworthiness, precision, and accuracy of claims made with the data. By using this methods based in uncertainty quantification, researchers may stop collecting data and/or human-coding as soon as sufficient trustworthiness is reached. 

\section{Conclusion}\label{Sec:conclusion}

We have presented a four-part methodology that applies established scientific tools and procedures (evaluating models, calculating statistical uncertainty, calculating systematic uncertainty) to build trust in effective machine coding. We demonstrated this method is effective for more than one coding scheme and provided a real-world example of using our method to machine code data from a new data set without additional human coding. We propose (and provide evidence for) several recommended best practices in machine coding: balancing training sets across code outcomes, 
drawing comparisons between machine coding mechanisms and human coding mechanisms (coding schemes), and measuring and accounting for statistical and systematic uncertainty whenever a quantitative claim is made. We hope these recommendations can continue to evolve as the use of machine learning and natural language processing is applied to PER data.

\acknowledgments{We thank Matthew Dew, Lauren Douglass, Daniella Garcia Almeida, Sophia Jeon, and Ali Mazrui for assistance with hand-coding. Thank you to Nicholas T Young for his data science expertise and detailed feedback, and thank you to Tor Ole Odden, Meagan Sundstrom, Marcos D Caballero, Carl Wieman, and Megan Flynn for insightful comments on early versions of this article. This work was supported by NSF grants DUE-1808945 and DUE-2000739.}

\bibliography{bibliography.bib} 

\appendix

\section{Effect of training set imbalance}\label{Sec:app_training}

\subsection{Training set balance for the Limitations code}

In the main text, we presented evidence with the CR code in the Trustworthy data set that balanced training sets (i.e., 50\% of the responses include the code) are most effective for training a model. We additionally assess this effect for the L code in the Sources data set with the same procedure. We train our algorithm with nine different training sets, each with a different value of $p_{train}$ between 0.1 and 0.9 inclusive. The size of the training set for each algorithm was 600. For each algorithm, we compute $E_C$ and $E_H$ by pulling 100 sample test sets of size $n=100$ from a test bank of size $N_{bank}=200$. We fix the proportion of responses in the test bank containing the code to 0.5.

\begin{figure}{\includegraphics[width=0.9\linewidth]{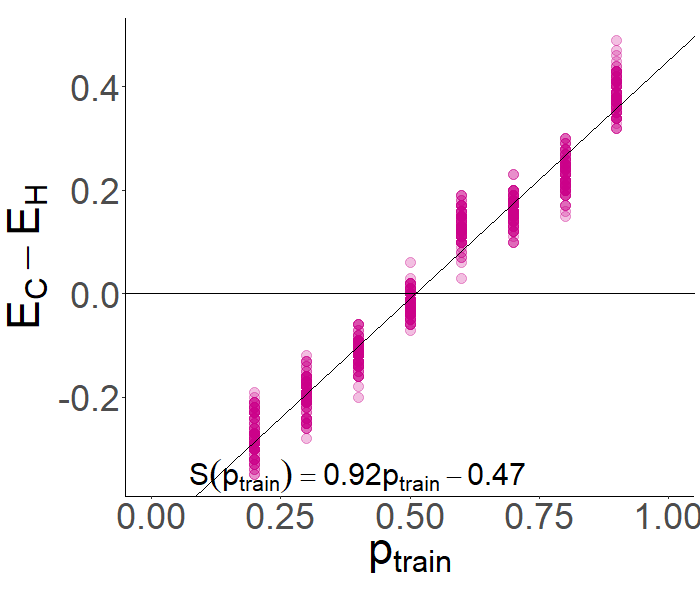}}
  \caption{Systematic uncertainty increases as $p_{train}$ is less balanced ($p_{train} = 0.5$). (Code: Limitations, proportion of responses containing the code in the test bank is 0.5).~\label{fix_test_vary_train_L}}\vspace{-1em}
\end{figure}

Fig.~\ref{fix_test_vary_train_L} shows that the difference between $E_C$ and $E_H$ is zero for $p_{train}=0.5$ but deviates from zero as $p_{train}$ deviates from 0.5. Given that this finding is consistent for two different codes from distinct data sets, we suggest the finding is likely generalizable to other codes as well, though future work should continue to test this empirically.


\begin{figure*}
  \subfloat[]{\includegraphics[width=0.45\linewidth]{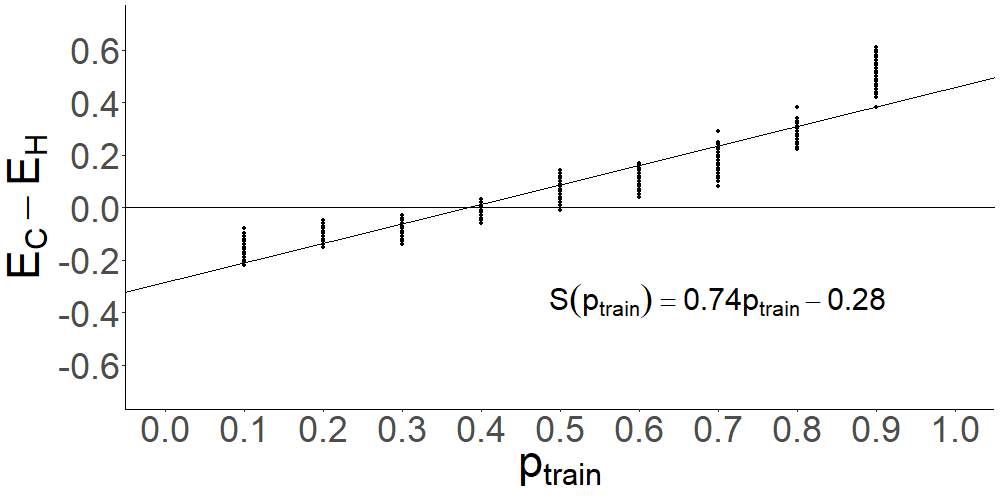}}
 \subfloat[]{\includegraphics[width=0.45\linewidth]{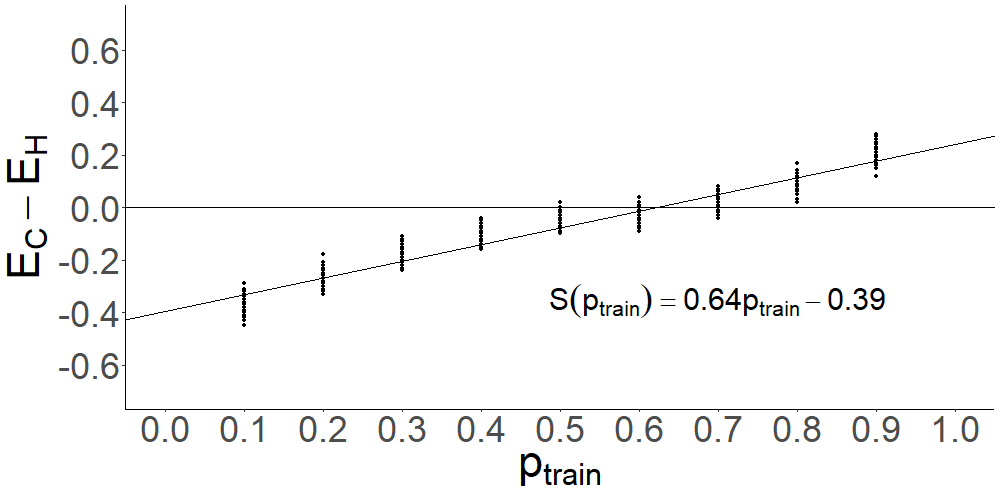}}
  \caption{a) $S(p_{train})$ where the proportion of responses in the test bank that contain the code (as determined by human coders) is 0.3, b) $S(p_{train})$ where the proportion of responses in the test bank that contain the code (as determined by human coders) is 0.7.~\label{appendix_other_test_prop}}\vspace{-1em}
\end{figure*}

\subsection{Training set balance for varying proportions of responses in the test bank that contain the code}

In the main text, we demonstrated the benefit of balancing a training set using a test bank where the proportion of responses in the test bank containing the code is also balanced (50\% of the responses contain the code). In this section, we demonstrate what happens at other proportions of responses in the test bank.



We test values of $p_{train}$ between 0.1 and 0.9 (in steps of 0.1) for two different test sets with different proportions of responses that contain the code: (a) $0.3$ and (b) $0.7$. We use the same algorithm for the CR code used above with a training set of size $600$. For each value of $p_{train}$, we take 100 sample test sets of size $n = 100$ and estimate $S(E_C)$ across the 100 samples. These data are displayed in Fig.~\ref{appendix_other_test_prop}.

We find that for the test bank where the proportion of responses that contain the code is $0.3$, $S(p_{train})$ is zero around $p_{train} = 0.4$. For the test bank where the proportion of responses that contain the code is $0.7$, $S(p_{train})$ is zero around $p_{train} = 0.6$. Thus, the optimal value of $p_{train}$ has some correlation with the proportion of responses in the test bank that contain the code.

In spite of the correlation between the optimal value of $p_{train}$ and the proportion of responses in the test bank that contain the code, we argue that a balanced training set with $p_{train}=0.5$ is the best choice for training a model. First, in both of our test banks in Fig.~\ref{appendix_other_test_prop}, the systematic is within statistical uncertainty to zero when $p_{train} = 0.5$. Second, when constructing a training set, we do not know the proportion of responses in the new data sets to which the model will be applied, and thus cannot make \textit{a priori} decisions about the best proportion to use in the training set based on the correlation with the new data set. Thus, we must determine what value of $p_{train}$ is optimal \emph{on average} across all possible proportions of responses that contain the code in the test set. Based on the analyses here, we recommend balancing the training set (fix $p_{train} = 0.5$) and modeling the remaining systematic using $S(E_C)$ (see Sec.~\ref{systematics}).

\section{Statistical Uncertainty}\label{Sec:app_statistical}

\subsection{Validity of statistical uncertainty expression for the L code}

In the main text, we present evidence with the CR code that Eqn.~\ref{eqn:sigma_EC_stats} describes the statistical variability of the machine coded responses. Here we assess the applicability of Eqn.~\ref{eqn:sigma_EC_stats} to the L code using a training set of size 600 with $p_{train}=0.5$, using the same procedures as with the CR code in the main text. 

\begin{figure*}
  \subfloat[]{\includegraphics[width=0.45\linewidth]{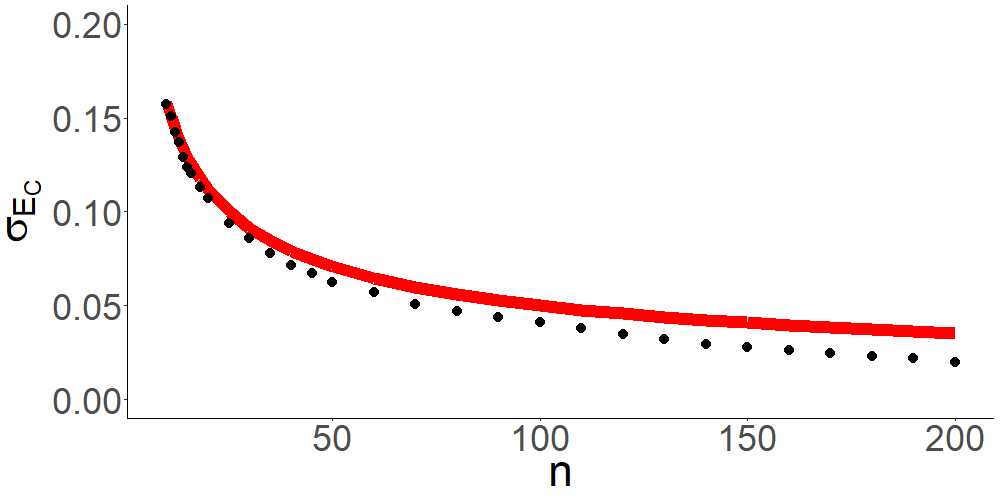}}
 \subfloat[]{\includegraphics[width=0.45\linewidth]{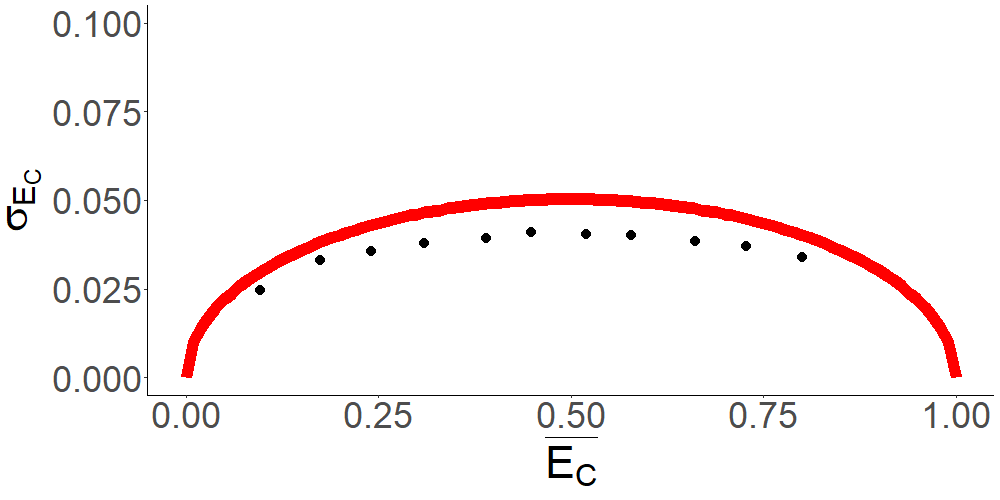}}
  \caption{Comparison between empirical data (black points) and Eqn. \ref{eqn:sigma_EC_stats} (red lines). a) Standard deviation of $E_C$ as $n$, the number of responses in each sample test set, varies empirically (code: L, $p_{train} = 0.5$, the proportion of responses in the test bank that contain the code is 0.5 and $\overline{E_C}\approx 0.5$).  b) Standard deviation of $E_C$ as $\overline{E_C}$ varies empirically (code: L, $n = 100$, $p_{train} = 0.5$).~\label{sigma_E_C_L}}\vspace{-1em}
\end{figure*}

We first probe the relationship between $\sigma _{E_C}$ and $n$ by testing 30 different values of $n$ between 10 and 200. For each value of $n$, we draw 1000 test set samples of size $n$ from a test bank where the proportion of responses containing the code is 0.5. We then have our algorithm compute $E_C$ for every sample. For each value of $n$, we calculate an empirical value for $\sigma _{E_C}$ by computing the standard deviation across the 1000 samples. As for the CR code, we find that the empirical values of $\sigma _{E_C}$ approximately follow the predicted $1/\sqrt{n}$ relationship but that Eqn.~\ref{eqn:sigma_EC_stats} slightly overestimates $\sigma _{E_C}$ (Fig.~\ref{sigma_E_C_L}, left).

We then probe the relationship between $\sigma_{E_C}$ and $\overline{E_C}$ for the L code. We test values of $\overline{E_C}$ between 0 and 1. We create a set of test banks where the proportion of responses containing the code ranges from 0 to 1 inclusive in steps of 0.1. We draw repeated samples from each test bank to calculate $\overline{E_C}$ for each test bank. For each value of $\overline{E_C}$, we take 1000 sample test sets of size $n=100$ and calculate $\sigma _{E_C}$ across the 1000 samples. As for the CR code, we find that the empirical values of $\sigma _{E_C}$ approximately follow the predicted parabolic relationship but that Eqn.~\ref{eqn:sigma_EC_stats} slightly overestimates $\sigma _{E_C}$ (Fig.~\ref{sigma_E_C_L}, right).

\subsection{Validity of the estimate \texorpdfstring{$\sqrt{\frac{\overline{E_C}(1-\overline{E_C})}{n}} \approx \sqrt{\frac{E_C(1-E_C)}{n}}$}{uncertainty with mean of EC is approximately uncertainty with single measurement of EC}}

Here we assess the quality of the estimate $\sqrt{\frac{\overline{E_C}(1-\overline{E_C})}{n}} \approx \sqrt{\frac{E_C(1-E_C)}{n}}$ that is made in Eqn.~\ref{eqn:sigma_EC_stats}. We check the assumption as a function of $n$ and as a function of $\overline{E_C}$.

We train an algorithm for the CR code using a training set of size $600$ with $p_{train} = 0.5$. We use 1000 test set samples for each value of $n$ and $\overline{E_C}$, such that each test set sample can be used to make an individual measurement $E_C$. We compute $\sigma_{E_C}\approx\sqrt{\frac{E_C(1-E_C)}{n}}$ for each individual $E_C$ value from each test set sample. For each value of $n$ and $\overline{E_C}$ we compute the $95\%$ confidence interval for the 1000 values and display it in blue to demonstrate the deviation from $\sqrt{\frac{\overline{E_C}(1-\overline{E_C})}{n}}$ (in red). 

\begin{figure*}
  \subfloat[]{\includegraphics[width=0.45\linewidth]{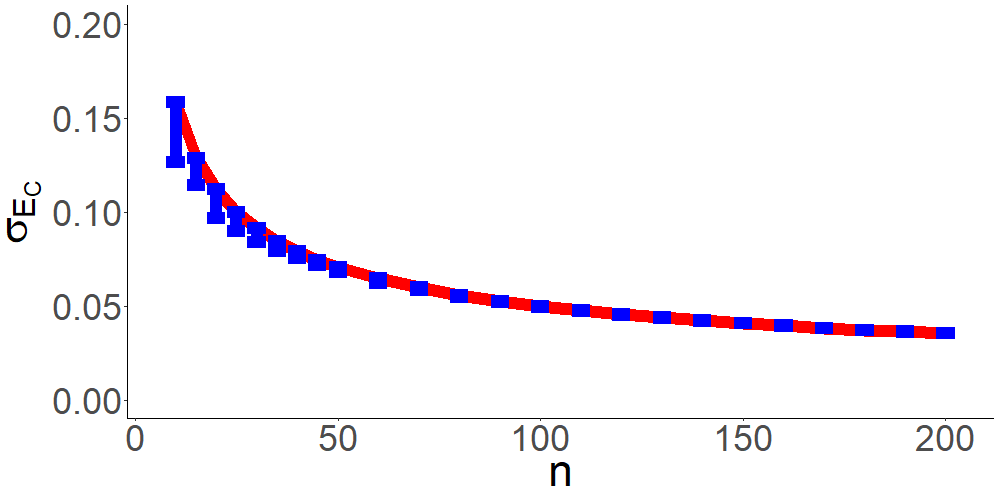}}
 \subfloat[]{\includegraphics[width=0.45\linewidth]{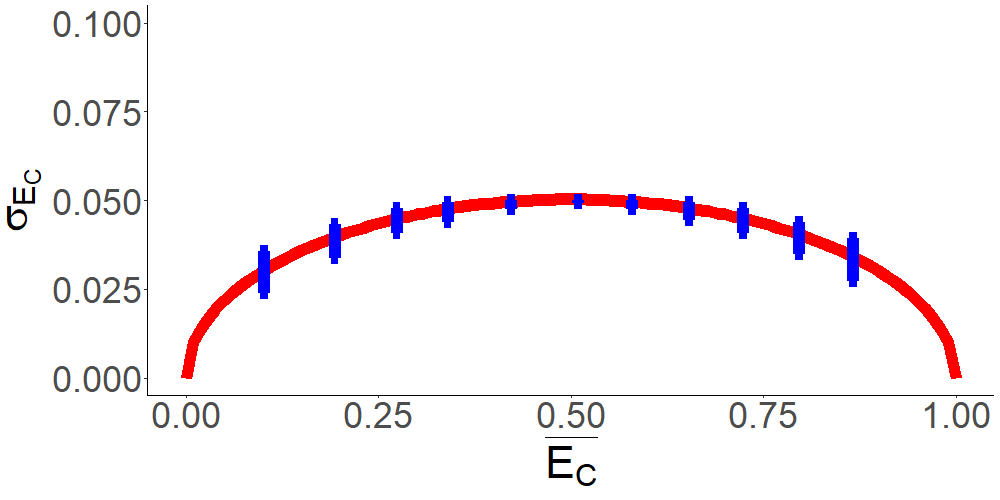}}
  \caption{Accuracy of the approximation $\sqrt{\frac{\overline{E_C}(1-\overline{E_C})}{n}} \approx \sqrt{\frac{(E_C(1-E_C)}{n}}$. Blue errorbars show 95\% interval of the approximation out of the 1000 samples.~\label{approximation}}\vspace{-1em}
\end{figure*}

To evaluate the quality of the estimate as $n$ varies, we test samples of size $n$, where $n$ varies between 10 and 200. For each value of $n$, we draw 1000 test set samples of size $n$ from a test bank with 50\% of the responses containing the code. This corresponds to $\overline{E_C}\approx 0.5$. We compute $E_C$ and $\sqrt{\frac{E_C(1-E_C)}{n}}$ for each test set sample and then calculate $\sqrt{\frac{\overline{E_C}(1-\overline{E_C})}{n}}$ at each value of $n$. The results are plotted in Fig.~\ref{approximation}a. Although there is moderately large deviation for $n<30$, for most values of $n$ the size of the difference is equal to or less than a rounding error if the statistical uncertainty is reported with 1-2 digits.

We similarly evaluate the quality of the estimate as $\overline{E_C}$ varies. We test values of $\overline{E_C}$ between 0 and 1. From the test data, we create 11 different test banks where the proportion of responses containing the code (as determined by the human coder) ranges from 0 to 1 inclusive in steps of 0.1. For each value of $\overline{E_C}$ (each test bank), we draw 1000 test set samples of size $n = 1000$. We compute $E_C$ and $\sqrt{\frac{E_C(1-E_C)}{n}}$ for each test set sample and then calculate $\sqrt{\frac{\overline{E_C}(1-\overline{E_C})}{n}}$ at each value of $\overline{E_C}$. The results are plotted in Fig.~\ref{approximation}b. For most values of $\overline{E_C}$ included in the plot, the size of the difference is equal to or less than a rounding error if the statistical uncertainty is reported with 1-2 digits. There are a few areas where the deviation is a bit larger, in particular where $\overline{E_C} < 0.2$ or $\overline{E_C} > 0.8$. 


\subsection{Independence of statistical uncertainty with training set size}

Here we demonstrate that the statistical uncertainty $\sigma _{E_C}$ is independent of the training set size. We separately train algorithms for the CR code and the L code using $p_{train}=0.5$ and training set sizes $N_{train}$ ranging from 100 to 600 inclusive in steps of 50. For each value of $N_{train}$, we draw 100 test set samples of size $n=100$ from a test bank of 200 responses with the proportion of responses in the test bank that contain the code fixed at 0.5. We then calculate $\sigma _{E_C}$ for each value of $N_{train}$.

\begin{figure*}
  \subfloat[]{\includegraphics[width=0.45\linewidth]{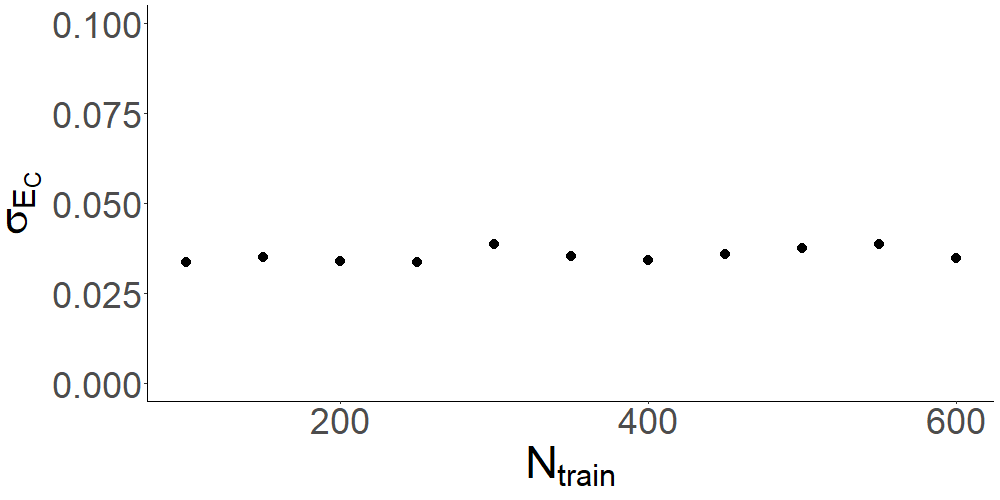}}
 \subfloat[]{\includegraphics[width=0.45\linewidth]{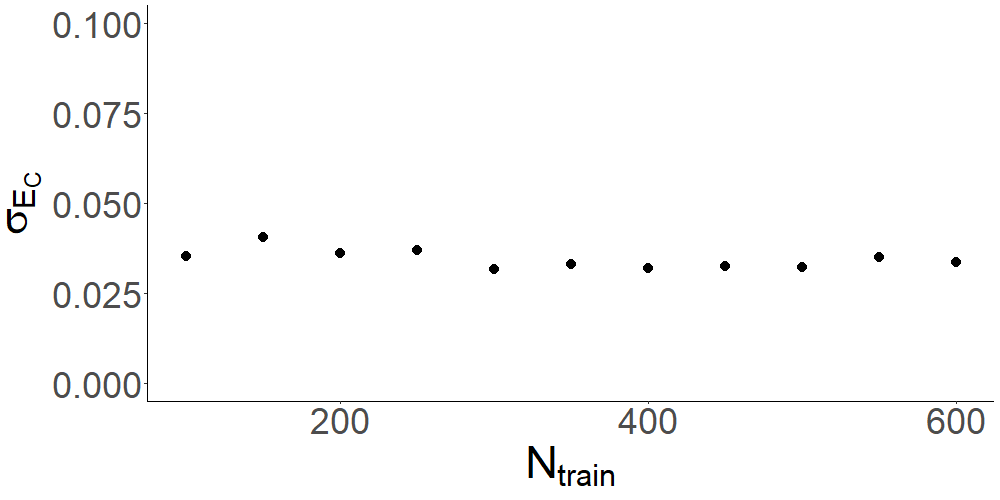}}
  \caption{Statistical uncertainty is independent of the size of the training set. a) Consistent Results code. b) Limitations code. $n = 100, p_{train} = 0.5$, the proportion of responses in the test bank that contain the code is $0.5$, $N_{bank} = 200$~\label{sigma_E_C_vs_train_size}}\vspace{-1em}
\end{figure*}

Fig. \ref{sigma_E_C_vs_train_size} demonstrates that $\sigma_{E_C}$ does not vary based on $N_{train}$ for either the (a) CR code or (b) L code. This result provides evidence that statistical uncertainty is an effect of the limited sample size rather than the size of the training set (a proxy for the quality of the trained model). 

\section{Effect of different training sets on models of systematic uncertainty}\label{Sec:app_systematic}

\begin{figure*}
  \subfloat[]{\includegraphics[width=0.45\linewidth]{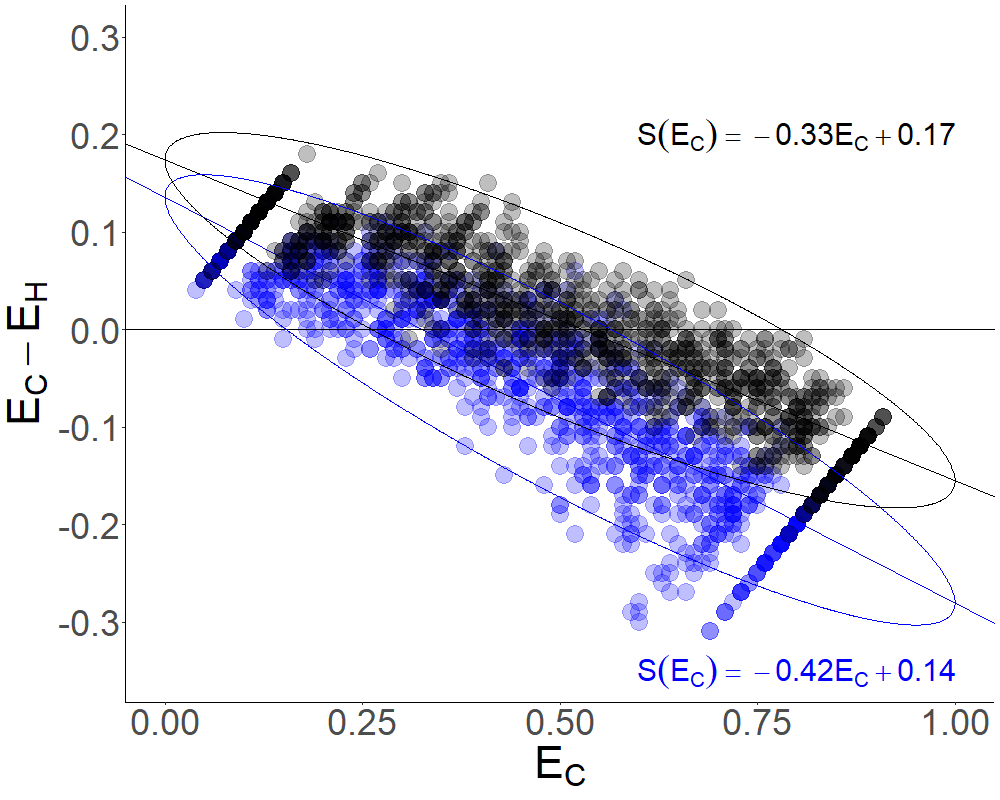}}
 \subfloat[]{\includegraphics[width=0.45\linewidth]{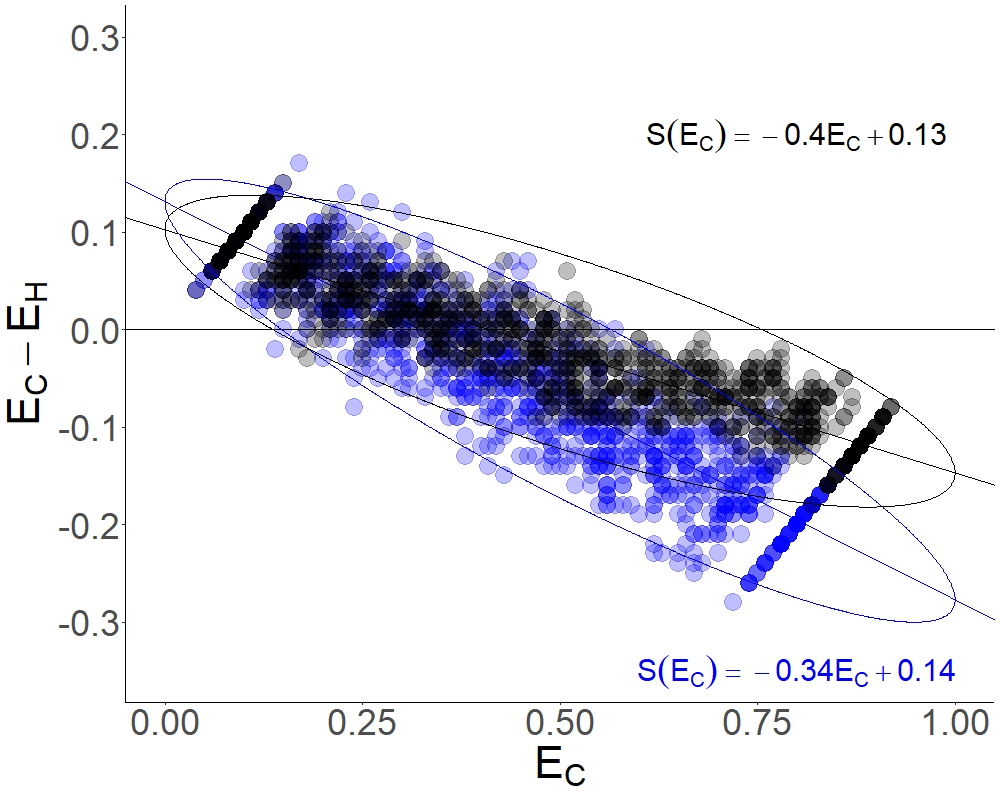}}
  \caption{Given a value of $E_C$, we can compute the systematic uncertainty using the line of best fit $S(E_C)$. $S(E_C)$ is specific to a training set. When the training data changes, with all else staying constant, $S(E_C)$ changes. Scatter around line of best fit is a result of statistical uncertainty as less than 95\% of the data falls within the black oval depicting $2*\sigma_{E_C}$ (see Eqn.\ref{eqn:sigma_EC_stats}. a) $S(E_C)$ for CR code with a different training set (blue) along with $S(E_C)$ from Fig.~\ref{systematics} left panel (black). (Code: CR, $p_{train} = 0.5$) b) $S(E_C)$ for L code with a different training set (blue) along with $S(E_C)$ from Fig.~\ref{systematics} right panel (black) (Code: L, $p_{train} = 0.5$).~\label{systematics_other_training}}\vspace{-1em}
\end{figure*}

In the main text, we provide evidence that the systematic difference between human and machine coders can be modeled as a linear function $S(E_C)$. Here, we provide evidence that the specific function is dependent on the trained model. That is, if the training data used to generate a trained model are changed, $S(E_C)$ changes. This analysis informs our recommendation to generate $S(E_C)$ for \textit{a single trained model} that will be applied to new data.  

We sample a training set of size 600 for the CR code (different from the training set used in Fig.~\ref{systematics}), both of which are still balanced (50\% of responses contain the code). From the data not included in this training set, we generate a set of nine different test banks where the proportion of responses containing the code varies from $0.1$ to $0.9$ inclusive. The size of each test bank is $N_{bank} = 300$. For each test bank, we compute $E_C$ and $E_H$ for 100 sample test sets of size $n = 100$ pulled from the test bank. In Fig.~\ref{systematics_other_training}a), we plot $E_C - E_H$ vs. $E_C$ for this new data (in blue) alongside the data from Fig.~\ref{systematics} (in black). We repeat this same process for the L code (panel b). 

For the CR code, the difference between systematics for the two different trained models extends beyond statistical uncertainty throughout the range of $E_C$. For instance, for the CR code $S(E_C = 0.25) = 0.088$ for the original trained model in Fig.~\ref{systematics}, but $S(E_C = 0.25) = 0.035$ when using the different training set shown in Fig.~\ref{systematics_other_training}. The statistical uncertainty where $E_C=0.25$ is $\sigma_{E_C}(0.25)=0.043$; thus, the difference in these two estimate, $0.088-0.035=0.053$, is slightly larger than $\sigma_{E_C}(0.25)$. The difference grows as $E_C$ increases. When $E_C=0.75$, for example, $S(E_C = 0.75) = -0.077$ for the original trained model and $S(E_C = 0.75) = -0.175$ for the different training set. This difference, $-0.077-(-0.175)=0.098$, is more than twice the statistical uncertainty $\sigma _{E_C}(0.75)=0.043$.

For the L code, the difference between systematics for two different trained models is indistinguishable for $E_C < 0.5$ but extends beyond the statistical uncertainty at high values of $E_C$. Thus, changing the trained model may be beyond statistical uncertainty, especially at extreme values of $E_C$. We expect that this occurs because each trained model learns slightly different coding rules, even though both training sets are balanced. 

We note that in this analysis we are comparing only two specific examples of trained models for each code, thus the comparison above cannot be generalized to other examples. For instance, we cannot conclude that a third example of a trained model for the L code is more likely to be indistinguishable from our existing examples than a third example of a trained model for the CR code.

\end{document}